\documentclass[fleqn,10pt]{wlscirep}
\usepackage[utf8]{inputenc}
\usepackage[T1]{fontenc}
\usepackage{graphicx}
\usepackage{placeins}
\usepackage{float}
\usepackage{xcolor}
\usepackage{subfigure}
\usepackage{amsmath}
\usepackage{comment}
\usepackage{tikz}
\usepackage{stackengine}
\newcommand\ubar[1]{\stackunder[1.2pt]{$#1$}{\rule{.8ex}{.075ex}}}

\usetikzlibrary{shapes, arrows}
\tikzstyle{decision 1} = [diamond, draw, fill=blue!20, 
    text width=3.5em, text badly centered, node distance=1.5cm, inner sep=0pt]
\tikzstyle{decision 2} = [diamond, draw, fill=red!20, 
    text width=3.5em, text badly centered, node distance=1.5cm, inner sep=0pt]
\tikzstyle{decision 3} = [diamond, draw, fill=green!20, 
    text width=3.5em, text badly centered, node distance=1.5cm, inner sep=0pt]
\tikzstyle{block 1} = [rectangle, draw, fill=blue!20, 
    text width=6.5em, text centered, rounded corners, minimum height=1em]
\tikzstyle{block 2} = [rectangle, draw, fill=red!20, 
    text width=7em, text centered, rounded corners, minimum height=1em]
\tikzstyle{block 3} = [rectangle, draw, fill=green!20, 
    text width=7em, text centered, rounded corners, minimum height=1em]
\tikzstyle{long block 0} = [rectangle, draw, fill=orange!20, 
    text width=16em, text centered, rounded corners, minimum height=1em]
\tikzstyle{long block 1} = [rectangle, draw, fill=blue!20, 
    text width=12em, text centered, rounded corners, minimum height=1em]
\tikzstyle{long block 2} = [rectangle, draw, fill=red!20, 
    text width=12em, text centered, rounded corners, minimum height=1em]
\tikzstyle{long block 3} = [rectangle, draw, fill=green!20, 
    text width=10em, text centered, rounded corners, minimum height=1em]
\tikzstyle{block 4} = [rectangle, draw, fill=yellow!20, 
    text width=25em, text centered, rounded corners, minimum height=1em]
\tikzstyle{line} = [draw, -latex']

\title{The phase unwrapping of under-sampled interferograms using radial basis function neural networks }

\author[1,*]{P.-A. Gourdain}
\author[1]{A. Bachmann}
\affil[1]{Physics and Astronomy Department,University of Rochester,Rochester NY14627, USA}

\affil[*]{gourdain@pas.rochester.edu}


\begin{abstract}
Interferometry can measure the shape or the material density of a system that could not be measured otherwise by recording the difference between the phase change of a signal and a reference phase.  This difference is always between $-\pi$ and $\pi$ while it is the absolute phase that is required to get a true measurement. There is a long history of methods designed to recover accurately this phase from the phase "wrapped" inside $]-\pi,\pi]$. However, noise and under-sampling limit the effectiveness of most techniques and require highly sophisticated algorithms that can process imperfect measurements. Ultimately, analysing successfully an interferogram amounts to pattern recognition, a task where radial basis function neural networks truly excel at. The proposed neural network is designed to unwrap the phase from two-dimensional interferograms, where aliasing, stemming from under-resolved regions, and noise levels are significant. The neural network can be trained in parallel and in three stages, using gradient-based supervised learning. Parallelism allows to handle relatively large data sets, but requires a supplemental step to synchronized the fully unwrapped phase across the different networks. 
\end{abstract}
\begin{document}

\flushbottom
\maketitle
\thispagestyle{empty}

\section*{Introduction}
Interferometry has been used successfully in measuring quantities that would be otherwise unpractical or difficult to measure\cite{Oppenheim81}. However, the absolute phase $\phi_{GT}$, or "ground truth" cannot be measured directly. Rather, only its wrapped value\cite{Blackledge89} $\phi_W$, bounded between  $-\pi$ and $\pi$, can be computed from the interferogram. As the wrapped phase $\phi_W$ cannot be used directly, the data needs to be unwrapped, a task deceptively challenging, especially in the presence of noise. To further complicate matters, large chunks of phase data might be missing due to the presence of strong signal cut-off (often seen in magnetic resonance imaging\cite{WITOSZYNSKYJ2009257}) or under-sampling, as in dense plasma interferometry\cite{Sarkisov96}. Further, data sets have grown extremely large, straining serial algorithms used in phase unwrapping (e.g functional MRI\cite{STANLEY2021117631}, interferometric synthetic aperture radar\cite{yu2019}, shape reconstruction\cite{Zhang:07} or fringe projection profilometry\cite{gorthi2010fringe}). Regardless of the problem, the phase unwrapping procedure needs to find an approximate phase $\phi$ that is such that $\phi=\phi_{GT}+\text{o}(\phi_{GT})$. The ground truth must be extracted from the intensity $I$ given by
\begin{equation}\label{eq:intensity}
    I(x,y)=A(x,y)+B(x,y)\cos^2\phi_{GT}(x,y).
\end{equation}
Note that in the ideal case, where $A\equiv1$ and $B\equiv1$, we effectively measure the wrapped phase $\phi_W=W(\phi_{GT})$, where $W$ is the wrapping operator defined as
\begin{equation}\label{eq:wrapping_operator}
W(\phi)=\phi-2k\pi\text{ with }k \in \mathbb Z \text{ such that } W(\phi) \in ]-\pi,\pi].    
\end{equation}
In this work we demodulate the phase, i.e. turning the intensity data given by Eq. \ref{eq:intensity} into wrapped phase, using filtered Fourier transforms\cite{Takeda:82,Macy:83,Roddier:87}. 

When the phase $\phi_{GT}$ is well-behaved (i.e. continuous, noise free, over-sampled) then phase unwrapping is straightforward \cite{itoh82}. However, when the phase is corrupted by noise or under-sampled (i.e. aliasing), unwrapping becomes difficult\cite{ghiglia98} and a variety of methods have been developed to overcome this issue. Early methods used branch-cuts\cite{cusack95,goldstein98,zheng11}, least-square algorithm\cite{Ghiglia:94,wang17} or polynomial phase approximation\cite{katkovnik08,gorthi09}. However, they also did not  react well to noise, leading to the development of algorithms capable of handling high noise levels\cite{servin98,loffeld07,xie14,xie15,Cheng:15,kulkarni18} using Kalman filters\cite{kalman60,julier04}. Machine learning algorithms were equally successful, using artificial neural networks\cite{schwar00}, then deep learning\cite{wang19,Yin2019,zhang19,qin20} and finally convolutional networks\cite{yang21,perera2021joint}. Unlike previous techniques, which tend to use the grid given by the natural data layout, machine learning is usually not relying on the physical data structure to preform the unwrapping. Compared to other phase-based measurements, high energy density plasma interferometry \cite{Domier88,Thaury13,Swadling14} has its set of unique challenges. Typically, external noise levels are relatively low since most interferometers use a laser beam\cite{lebedev02} that is extremely bright ($>100\text{ MW/cm}^2$). However, diffraction can generate artifacts that degrade beam quality. Further, these plasmas have energy densities on the order of $1\text{ kJ/cm}^3$, and the continuum light they produce as a result can cause large-scale intensity blotches embedded inside the interferogram. High energy density plasmas can also be surrounded by complex structures, which blocks part of the beam and create regions free of interference fringes. The shape of these structures is often complex\cite{ampleford08,hasson20} and required to be removed from the input data. Finally, with electron density gradients relatively large, interferometry data is often under-sampled, creating zones where the interference pattern is not directly usable\cite{gourdain13}. 

To deal with such practical considerations, we developed a parallel neural network algorithm capable of unwrapping phase data following a staged supervised learning. Parallelization is obtained by clustering neurons across contiguous regions, where overlapping neurons, called \textit{ghost} neurons\cite{long2008scalable}, are used to synchronized the different networks. Unlike previous methods using Levenberg-Marquardt algorithms \cite{Kulkarni17} or spatial derivatives\cite{Gao:08}, the proposed radial basis neural networks \cite{Broomhead:1988} (RBFNN) to analyze interferograms with under-sampled regions caused by inadequate digitization or lossy data compression.

\section*{Training of the staged neural network}
\subsection*{Preliminary remarks}
\subsubsection*{A condition to detect aliasing}
Super-resolution imaging uses numerical or physical techniques that allow to effectively increase the resolution of an image\cite{Wang21,Bizhani22}. For interferometry data, this technique is required when the phase was wrapped more than once across two pixels, a phenomenon known as aliasing\cite{ghiglia98, Kulkarni20}. It can present itself as a series of swift, consecutive jumps, a case relatively easy to detect when noise levels are low. It can also be completely inconspicuous. For instance, a phase which varies as $\phi_{GT}(n)=\xi-\pi$ and $\phi_{GT}(n+1)=\xi+\pi$, where $0<\xi\ll 1$, would yield a seemingly constant wrapped phase $\phi_W(n)=\xi-\pi$ and $\phi_W(n+1)=\xi-\pi$.
As a lower bound, we can see that aliasing is present when $|\partial\phi_{GT}|>2\pi$, then $W(\partial\phi_{GT})$ has jumps inside $]-\pi,\pi]$. While this is only a necessary condition, it becomes sufficient when $\phi_{GT}$ is continuous, as we will see later, allowing to detect the presence of aliasing inside the data.
\subsubsection*{A key property of periodic functions}
When any phase $\Phi$ is discretized, we can define its left derivative as $\partial_L\Phi(n)=\Phi(n)-\Phi(n-1)$. Using Eq. \ref{eq:wrapping_operator}, we get  $\mathcal{F}\big(\partial_L\phi_{GT}(n)\big)=\mathcal{F}\big(\phi_W(n)-\phi_W(n-1)+2(k_n-k_{n-1})\pi\big)$ for any $2\pi$-periodic function $\mathcal{F}$. As a result, $\mathcal{F}\left(\partial_L\phi_{GT}(n)\right)=\mathcal{F}\left(\partial_L\phi_{W}(n)\right)$. This is true for the whole domain if we use the linear extrapolation to define the left derivative at the left boundary as $\partial_L\Phi(1)=-\Phi(1)+2\Phi(2)-\Phi(3)$. 
For the right derivative, defined a $\partial_R\Phi(n)=\Phi(n+1)-\Phi(n)$, we use the same reasoning to get $\mathcal{F}\big(\partial_R\phi_{GT}(n)\big)=\mathcal{F}\big(\partial_R\phi_{W}(n)\big)$. This is valid across the whole domain if we now used the linear extrapolation of the right derivative at the right boundary as $\partial_R\Phi(N)=-\Phi(N-2)+2\Phi(N-1)-\Phi(N)$.  This property does not extend to the central derivative $\partial_C\Phi(n)=\frac{1}{2}[\Phi(n+1)-\Phi(n-1)]$ since $\mathcal{F}\big(\partial_C\phi_{GT}(n)\big)=\mathcal{F}\big(\partial_C\phi_W(n)+(k_{n+1}-k_{n-1})\pi\big)$ is equal to $\mathcal{F}\big(\partial_C\phi_W(n)\big)$ only when $k_n-k_{n-1}$ is even, but not when it is odd. In the end, we find
\begin{equation}\label{eq:periodic_derivative}
    \mathcal{F}(\partial_{L,R}\phi_W)=\mathcal{F}(\partial_{L,R}\phi_{GT}).
\end{equation}
\noindent It happens that the second derivative $\partial^2\Phi=\Phi(n+1)-2\Phi(n)+\Phi(n-1)$ is also invariant since $\mathcal{F}\big(\partial^2\phi_{GT}(n)\big)=\mathcal{F}\big(\phi_W(n+1)-2\phi_W(n)+\phi_W(n-1)+2(k_{n+1}-2k_n+k_{n-1})\pi\big)=\mathcal{F}\big(\partial^2\phi_W(n)\big)$. If we linearly extrapolate the second derivatives to the domain boundaries as $\partial^2\phi(1)=\phi(1)-3\phi(2)+3\phi(3)-\phi(4)$ and $\partial^2\phi(N)=\phi(N-3)-3\phi(N-2)+3\phi(N-1)-\phi(N)$, then this result is still valid for the whole domain and we get 
\begin{equation}\label{eq:periodic_2_derivative}
    \mathcal{F}(\partial^2\phi_W)=\mathcal{F}(\partial^2\phi_{GT}).
\end{equation}
Applying the same reasoning while using Eqs. \ref{eq:periodic_derivative} and \ref{eq:periodic_2_derivative} we can easily show that 
\begin{equation}\label{eq:periodic_wrapping_derivative}
    W(\partial_{L,R}\phi_W)=W(\partial_{L,R}\phi_{GT}),
\end{equation}
also known as the Itoh condition\cite{itoh82}, and
\begin{equation}\label{eq:periodic_wrapping_2_derivative}
    W(\partial^2\phi_W)=W(\partial^2\phi_{GT}).
\end{equation}

\subsubsection*{A restriction on the ground-truth}
An important restriction arises when unwrapping the phase using RBFNNs. Since the output layer is a sum of radial basis functions, which are smooth, the output is also smooth. So, the RBFNN can only unwrap a phase which ground truth is smooth. However, when few discontinuities are present, they can be hidden relatively easily from the RBFNN using a mask. This condition is usually not restrictive for interferograms generated by high energy density plasmas and it brings with it an essential component to a successful training. For instance, if we work with a phase $\phi_{GT}$ that is twice-continuous and not aliased, i.e. $\partial_{L,R}\phi_{GT}\in]-\pi,\pi]$, Eq. \ref{eq:periodic_wrapping_derivative} gives
\begin{equation*}
    W(\partial_{L,R}\phi_W)=\partial_{L,R}\phi_{GT}.
\end{equation*}
As a result, $W(\partial_{L,R}\phi_W)$ is continuous since $\partial_{L,R}\phi_{GT}$ is continuous, regardless of how many phase jumps are present in $\partial_{L,R}\phi_W$\cite{itoh82}. 

Since our goal is to deal with aliased phase, we can use the much less restrictive assumption $\partial^2\phi_{GT}\in]-\pi,\pi]$, and Eq. \ref{eq:periodic_wrapping_2_derivative} gives
\begin{equation}\label{eq:wrapping_2_derivative}
    W(\partial^2\phi_W)=\partial^2\phi_{GT}
\end{equation}
Further, if $\partial^2\phi_{GT}$ is continuous across the whole domain then $W(\partial^2\phi_W)$ is also continuous everywhere. We will make both assumptions in the rest of the paper.

\subsection*{Construction of the input layer}

While $\phi_W$ has jumps, we have shown that $W(\partial_{L,R}\phi_W)$ and $W(\partial^2\phi_W)$ are continuous if $\phi_{GT}$ is twice continuous. Yet, we cannot match the RBFNN output $\phi$ to $\phi_{GT}$ using gradient-based optimization since the wrapping operator $W$, which turns $\phi_{GT}$ into $\phi_W$, is not differentiable. While gradient-free methods\cite{Datta16,Muller2020,kirkpatrick83,Rere15,gudise03,carvalho06} have been used successfully in machine learning, gradient-based methods are always preferred when available. Eqs. \ref{eq:periodic_derivative} and \ref{eq:periodic_2_derivative} shows that we can use the differentiable sine and cosine functions instead of $W$ where differentiability is required. As long as $\phi_{GT}$ is twice continuous, these functions remove the spurious discontinuities otherwise present in $\partial_{L,R}\phi_W$ and $\partial^2\phi_W$ at every phase jump of $\phi_W$.  

\subsubsection*{Input layer to achieve super-resolution}

We can now construct an input layer $i_{1,\dots,12}$, where all the data is continuous. At every location inside the interferogram, we get:
\begin{equation}\label{eq:inlayer}
    \begin{matrix}
    &i_3=\cos(\partial_{xL}\phi_W)&i_7=\cos(\partial_{xR}\phi_W)\\
    i_1=\cos(\phi_W)&i_4=\sin(\partial_{xL}\phi_W)&i_8=\sin(\partial_{xR}\phi_W)&i_{11}=W(\partial_{xx}\phi_W)\\
    i_2=\sin(\phi_W)&i_5=\cos(\partial_{yL}\phi_W)&i_9=\cos(\partial_{yR}\phi_W)&i_{12}=W(\partial_{yy}\phi_W)\\
    &i_6=\sin(\partial_{yL}\phi_W)&i_{10}=\sin(\partial_{yR}\phi_W)\\
    \end{matrix}
\end{equation}
We can now compare the input layer with the RBFNN output $\phi$ using the following set of equations
\begin{equation}\label{eq:outcomp}
    \begin{matrix}
    &o_3=\cos(\partial_x\phi)&o_7=\cos(\partial_x\phi)\\
    o_1=\cos(\phi)&o_4=\sin(\partial_x\phi)&o_8=\sin(\partial_x\phi)&o_{11}=\partial_{xx}\phi\\
    o_2=\sin(\phi)&o_5=\cos(\partial_y\phi)&o_9=\cos(\partial_y\phi)&o_{12}=\partial_{yy}\phi\\
    &o_6=\sin(\partial_y\phi)&o_{10}=\sin(\partial_y\phi)\\
    \end{matrix}
\end{equation}
Note that, while $i_{11}$ and $i_{12}$ are still using the wrapping operator, this operator is not present in the equations used to compare the input layer and the RBFNN output because we restricted the second derivative to be between $-\pi$ and $\pi$. As the wrapping $W$ has no effect of the RBFNN output, it has completely disappeared from $o_{11}$ and $o_{12}$ and we now can take their derivatives. However, this operator is still required on the LHS of $i_{11}$ and $i_{12}$ to remove the phase jumps in $\partial^2\phi_W$. Also note that we have dropped the subscript $L$ and $R$ for the first derivatives of the output of the RBFNN, since it is a sum of analytical functions, which derivatives can be computed exactly.  
\subsubsection*{Input layer when super-resolution is not required}

The training can be substantially simplified when super-resolution is not required, i.e. $\partial\phi_{GT}\in]-\pi,\pi]$. In this case, we can replace Eq. \ref{eq:inlayer} with
\begin{equation}\label{eq:inlayer_short}
    \begin{matrix}
    \ubar i_1=\cos(\phi_W)&\ubar i_3=W(\partial_{xL}\phi_W)&\ubar i_5=W(\partial_{xR}\phi_W)\\
    \ubar i_2=\sin(\phi_W)&\ubar i_4=W(\partial_{yL}\phi_W)&\ubar i_6=W(\partial_{yR}\phi_W)\\
    \end{matrix}
\end{equation}
and Eq. \ref{eq:outcomp} with
\begin{equation}\label{eq:outcomp_short}
    \begin{matrix}
    \ubar o_1=\cos(\phi)&\ubar o_3=\partial_x\phi&\ubar o_5=\partial_x\phi\\
    \ubar o_2=\sin(\phi)&\ubar o_4=\partial_y\phi&\ubar o_6=\partial_y\phi.\\
    \end{matrix}
\end{equation}

\subsection*{The activation function}
Throughout this paper, the RBFNN will use a compact Wendland function\cite{wend95} as the activation function. Such functions can be constructed easily starting from
\begin{equation*}
\psi_{p,0}(r) = (1 - r)_{+}^{p} = \left\{ \begin{matrix}
(1 - r)^{p}  \\
0 \\
\end{matrix}\ \begin{matrix}
\text{ for }0 \leq r \leq 1 \\
\text{for }r > 1 \\
\end{matrix} \right.\ 
\end{equation*}
and using
\begin{equation*}
\psi_{p,q}(r) = \mathfrak{I}^{q}\psi_{p,0}\text{ for }0 \leq r \leq 1
\end{equation*}
to increase the function smoothness. Here $p,\ q \in \mathbb{N}$. The operator $\mathfrak{I}$
above is defined as $\mathfrak{I}f(r) = \int_{r}^{\infty}{f(t)tdt}\text{\ for\ }0 \leq r$.
Wendland functions are $C^{k}$ and can be computed analytically. They
yield a strictly positive definite matrix in $\mathbb{R}^d$, where
\emph{d}\textless{}\emph{p} and \emph{k}=2\emph{q}. The subscripts of $\psi_{p,q}$ will be dropped in the rest of the paper. Each neuron $(x_n,y_n)$ in our two dimensional dataset is activated using such radial basis functions\cite{Broomhead:1988,Buhmann2010}. In this paper, we use exclusively the Wendland function $\psi$ given by
\begin{equation}\label{eq:Wnedland_3_2}
\psi(r)=  \left\{ \begin{matrix}
\frac{1}{3} (1-r)^6 \left[(35 r+18)r+3\right]  \\
0 \\
\end{matrix}\ \begin{matrix}
\text{\ \ \ \ \ for\ }0 \leq r \leq 1 \\
\text{for\ }r > 1 \\
\end{matrix} \right.\ 
\end{equation}
obtained for $p=3$ and $q=2$

\subsection*{The output layer}
The output layer $\phi$ is expressed as a sum of radial basis functions $\psi_n(x,y)=\psi(r_n)$ centered on each neuron $n$ located at $(x_n,y_n)$ and scaled by the weight $w_n$. The output layer of a RBFNN with $N$ neurons is continuous and defined as
\begin{equation}\label{eq:rbf_form}
\phi(x,y)=\sum_{n=1}^N w_n(x,y)\psi \left(r_n(x,y)\right) \text{ with } r_n(x,y)=\sqrt{(x-x_n)^2\rho^2_{x_n}+(y-y_n)^2\rho^2_{y_n}},
\end{equation}
where $\rho_{x_n}$ and $\rho_{y_n}$ are the activation distance inverses for the $n^{th}$ neuron along the x- and y-directions respectively. The analytical expression of the Jacobian matrix is greatly simplified when using the inverse of the activation distance. 
As discussed later in the paper, we need to match five constraints to give the neural network super-resolution, i.e. $\phi_{GT}$, $\partial_x\phi_{GT}$, $\partial_y\phi_{GT}$, $\partial_{xx}\phi_{GT}$, $\partial_{yy}\phi_{GT}$. To match five constraints, we need to inject three degrees of freedom inside the weights as 
\begin{equation}
   w_n(x,y)=a_n+b_n(x-x_n)+c_n(y-y_n). 
\end{equation}
Together with $\rho_{x_n},\text{ and }\rho_{y_n}$, we now have five degrees of freedom per neuron. Note that the weights $w_n$ are now local linear\cite{Chen06,Nekoukar2009} in $x$ and $y$.

\subsection*{Definition of the objective function}
\subsubsection*{Objective function with super-resolution}

We can now define the objective function $F(\textbf e)$, used by the training process to minimize the error vector $\textbf e =[e_1,\hdots,e_N]^T$ for all neurons $n\in\{1,\hdots,N\}$
\begin{equation}
    F(\textbf e)=\textbf{e}^\text{T}\textbf{e}=\sum_{n=1}^N\sum_{j=1}^{12}e^2_{jn}\text{ with }e_{jn}=o_{jn}-i_{jn}
\end{equation}
The error $e_{jn}$ is the difference between the $j^{th}$ input layer value computed at the location $(x_n,y_n)$, i.e. $i_{jn}$, and the $j^{th}$ output layer value computed at the same location, i.e. $o_{jn}$.
Since the training tries to match both left and right derivatives, we expect the total error to remain high, even after full convergence, since the training won't be able to match the left and right derivatives simultaneously. Therefore, we can define the error $\mathbf{\hat e}$ to estimate when our network is fully trained
\begin{equation}
    \mathbf{\hat e}^\text{T}\mathbf{\hat e}=\sum_{n=1}^N\sum_{j=1}^{8}{\hat e}^2_{jn}
\end{equation}
where $\hat e_{jn}=\hat o_{jn}-\hat i_{jn}$ and 
$$
\begin{matrix}
    \hat i_1=i_1&\hat i_2=i_2&\hat i_3=\frac{1}{2}(i_3+i_5)&\hat i_4=\frac{1}{2}(i_4+i_6)&\hat i_5=\frac{1}{2}(i_5+i_9)&\hat i_6=\frac{1}{2}(i_6+i_{10})&\hat i_7=i_{11}&\hat i_8=i_{12}\\ 
    \hat o_1=o_1&\hat o_2=o_2&\hat o_3=o_3&\hat o_4=o_4&\hat o_5=o_5&\hat o_6=o_6&\hat o_7=o_{11}&\hat o_8=o_{12}\\
\end{matrix}
$$

\subsubsection*{Objective function without super-resolution}

The objective function for interferograms where super-resolution is not needed is defined as $\ubar F(\ubar {\textbf e})$ and should be used to minimize the error vector $\ubar{\textbf e} =[\ubar e_1,\hdots,\ubar e_N]^T$ for all neurons $n\in\{1,\hdots,N\}$
\begin{equation}
    \ubar F(\ubar{\textbf e})=\ubar{\textbf{e}}^\text{T}\ubar{\textbf{e}}=\sum_{n=1}^N\sum_{j=1}^{6}\ubar e^2_{jn}\text{ with }\ubar e_{jn}=\ubar o_{jn}-\ubar i_{jn}
\label{eq:objective_function_short}
\end{equation}
As we did earlier we can define the error $\ubar{\mathbf{\hat e}}$ to better assess the actual convergence error  
\begin{equation}\label{eq:convergence_error}
    \ubar{\mathbf{\hat e}}^\text{T}\ubar{\mathbf{\hat e}}=\sum_{n=1}^N\sum_{j=1}^{8}{\ubar{\hat e}}^2_{jn}
\end{equation}
where $\ubar{\hat e}_{jn}=\ubar{\hat o}_{jn}-\ubar{\hat i}_{jn}$ and 
$$
\begin{matrix}
    \ubar{\hat i}_1=\ubar{i}_1&\ubar{\hat i}_2=\ubar{i}_2&\ubar{\hat i}_3=\frac{1}{2}(\ubar i_3+\ubar i_5)&\ubar{\hat i}_4=\frac{1}{2}(\ubar i_4+\ubar i_6)\\ 
    \ubar{\hat o}_1=\ubar o_1&\ubar{\hat o}_2=\ubar o_2&\ubar{\hat o}_3=\ubar o_3&\ubar{\hat o}_4=\ubar o_4\\
\end{matrix}
$$\subsubsection*{Regularisation}
Simple Bayesian regularisation\cite{MacKay92}, or more complex variants such as using Markov chain Monte Carlo\cite{Sariev20}, have been proposed to avoid over-fitting noisy data and it is necessary in the presence of noise. 
\begin{equation}
    F_R(\textbf e)=\sum_{n=1}^N\sum_{j=1}^{12}e^2_{jn}+\Omega\sum_{n=1}^N\left(a_n^2+b_n^2+c_n^2+\rho_{x_n}^2+\rho_{y_n}^2\right).
\end{equation}
We found that $\alpha$ should be 1 during the first and second train stages since the noise has the largest impact on the second derivative of the wrapped phase. Regularization is typically not necessary during the last stage of the training and we can use $\Omega=0$.
When super-resolution is not needed we use 
\begin{equation}
    \ubar F_R(\ubar{\textbf e})=\sum_{n=1}^N\sum_{j=1}^{6}\ubar e^2_{jn}+\alpha\sum_{n=1}^N\left(a_n^2+b_n^2+c_n^2+\rho_{x_n}^2+\rho_{y_n}^2\right).
\end{equation}

\begin{figure}[ht]
\centering
\begin{tikzpicture}[node distance=1cm,auto]
    \node [block 1] (stage 1 start) {\scriptsize $W(\partial^2\phi_W)$};
    \node [block 1, below of=stage 1 start] (compute input 1) {\scriptsize $\forall n,\,\textbf{i}_n=[i_{11n},i_{12n}]^\text{T}$};
    \node [block 1, below of=compute input 1] (compute output 1) {\scriptsize $\forall n,\,\textbf{o}_n=[o_{11n},o_{12n}]^\text{T}$};
    \node [block 1, below of=compute output 1] (compute error 1) {\scriptsize $\textbf{e}=\textbf{o}-\textbf{i}$};
    \node [decision 1, below of=compute error 1] (converged 1) {\scriptsize $||\mathbf{\hat e}||<\epsilon_1$};
    \node [block 1, below of=converged 1,yshift=-.5 cm] (compute Jacobian matrix 1) {\scriptsize $\textbf{J}=\left[\partial_{a_q,b_q,c_q}o_p\right]$};
    \node [block 1, below of=compute Jacobian matrix 1,yshift=-0.5cm] (solve 1) {\scriptsize $d[a_q,b_q,c_q]^\text{T}=\left(\textbf{J}^\text{T}\textbf{J}+\lambda\textbf{I}\right)^{-1}\textbf{J}^\text{T}\textbf{e}$};
    \node [long block 1, left of=converged 1,xshift = -1 cm,yshift=-.25cm,rotate=90] (compute weights 1) {\scriptsize $ [a_q,b_q,c_q]^\text{T}-d[a_q,b_q,c_q]^\text{T}\rightarrow[a_q,b_q,c_q]^\text{T}$};
    \node [long block 0, left of=compute weights 1,xshift =- .5 cm,rotate=90] (initialize weights) {\scriptsize $ \forall q,\,a_q,b_q,c_q\in]-\pi/2,\pi/2],\,\rho_{xq}=\rho_0,\,\rho_{yq}=\rho_1$};
    \path [line] (stage 1 start) -- (compute input 1);
    \path [line] (initialize weights) -- node  {\scriptsize Initial} (compute weights 1);
    \draw [line] (compute input 1.east) -| ([xshift=0.25cm]compute input 1.east) |- (compute error 1.east);
    \path [line] (compute output 1) -- (compute error 1);
    \path [line] (compute error 1) -- (converged 1);
    \path [line] (compute Jacobian matrix 1) -- (solve 1);
    \path [line] (converged 1) -- node [near start] {\scriptsize no} (compute Jacobian matrix 1);
    \path [line] (compute weights 1) |- (compute output 1);
    \path [line] (solve 1) -| (compute weights 1);
    \node [block 2,right of =stage 1 start,xshift=3.5cm] (stage 2 start) {\scriptsize $W(\partial_{L,R}\phi_W)$, $W(\partial^2\phi_W)$};
    \node [block 2, below of= stage 2 start] (compute input 2) {\scriptsize $\forall n,\,\textbf{i}_n=[i_{3n},\hdots,i_{12n}]^\text{T}$};
    \node [block 2, below of=compute input 2] (compute output 2) {\scriptsize $\forall n,\,\textbf{o}_n=[o_{3n},\hdots,o_{12n}]^\text{T}$};
    \node [block 2, below of=compute output 2] (compute error 2) {\scriptsize $\textbf{e}=\textbf{o}-\textbf{i}$};
    \node [decision 2, below of=compute error 2] (converged 2) {\scriptsize $||\mathbf{\hat e}||<\epsilon_2$};
    \node [block 2, below of=converged 2,yshift=-.5 cm] (compute Jacobian matrix 2) {\scriptsize $\textbf{J}=\left[\partial_{a_q,b_q,c_q}o_p\right]$};
    \node [block 2, below of=compute Jacobian matrix 2,yshift=-0.5cm] (solve 2) {\scriptsize $d[a_q,b_q,c_q]^\text{T}=\left(\textbf{J}^\text{T}\textbf{J}+\lambda\textbf{I}\right)^{-1}\textbf{J}^\text{T}\textbf{e}$};
    \node [long block 2, left of=converged 2,xshift = -1.25 cm,yshift=-.25cm,rotate=90] (compute weights 2) {\scriptsize $ [a_q,b_q,c_q]^\text{T}-d[a_q,b_q,c_q]^\text{T}\rightarrow[a_q,b_q,c_q]^\text{T}$};

    \draw [line] (converged 1.east) -| node [near start] {\scriptsize yes}
                 ([xshift=-1.45cm]stage 2 start.west) -- (stage 2 start.west);
    \path [line] (stage 2 start) -- (compute input 2);
    \draw [line] (compute input 2.east) -| ([xshift=0.25cm]compute input 2.east) |- (compute error 2.east);
    \path [line] (compute output 2) -- (compute error 2);
    \path [line] (compute error 2) -- (converged 2);
    \path [line] (compute Jacobian matrix 2) -- (solve 2);
    \path [line] (converged 2) -- node [near start] {\scriptsize no} (compute Jacobian matrix 2);
    \path [line] (compute weights 2) |- (compute output 2);
    \path [line] (solve 2) -| (compute weights 2);
    \node [long block 3,right of =stage 2 start, xshift=4.05cm] (stage 3 start) {\scriptsize $\phi_W$,  $W(\partial_{L,R}\phi_W)$, $W(\partial^2\phi_W)$};
    \node [block 3, below of= stage 3 start] (compute input 3) {\scriptsize $\forall n,\,\textbf{i}_n=[i_{1n},\hdots,i_{12n}]^\text{T}$};
    \node [block 3, below of=compute input 3] (compute output 3) {\scriptsize $\forall n,\,\textbf{o}_n=[o_{1n},\hdots,o_{12n}]^\text{T}$};
    \node [block 3, below of=compute output 3] (compute error 3) {\scriptsize $\textbf{e}=\textbf{o}-\textbf{i}$};
    \node [decision 3, below of=compute error 3] (converged 3) {\scriptsize $||\mathbf{\hat e}||<\epsilon_3$};
    \node [block 3, below of=converged 3,yshift=-.5 cm] (compute Jacobian matrix 3) {\scriptsize $\textbf{J}=\left[\partial_{a_q,b_q,c_q,\rho_{xq},\rho_{yq}}o_p\right]$};
    \node [block 3, below of=compute Jacobian matrix 3,yshift=-0.5cm] (solve 3) {\scriptsize $d[a_q,b_q,c_q,\rho_{xq},\rho_{yq}]^\text{T}=\left(\textbf{J}^\text{T}\textbf{J}+\lambda\textbf{I}\right)^{-1}\textbf{J}^\text{T}\textbf{e}$};
    \node [long block 3, left of=converged 3,xshift = -1.25 cm,yshift=-.25cm,rotate=90] (compute weights 3) {\scriptsize $ [a_q,b_q,c_q,\rho_{xq},\rho_{yq}]^\text{T}-d[a_q,b_q,c_q,\rho_{xq},\rho_{yq}]^\text{T}\rightarrow[a_q,b_q,c_q,\rho_{xq},\rho_{yq}]^\text{T}$};
    \node [block 4, below of=solve 2] (done) {Radial Basis Function Neural Network Trained};

    \draw [line] (converged 2.east) -| node [near start] {\scriptsize yes}
                 ([xshift=-1.3cm]stage 3 start.west) -- (stage 3 start.west);
    \path [line] (stage 3 start) -- (compute input 3);
    \draw [line] (compute input 3.east) -| ([xshift=0.25cm]compute input 3.east) |- (compute error 3.east);
    \path [line] (compute output 3) -- (compute error 3);
    \path [line] (compute error 3) -- (converged 3);
    \path [line] (compute Jacobian matrix 3) -- (solve 3);
    \path [line] (converged 3) -- node [near start] {\scriptsize no} (compute Jacobian matrix 3);
    \path [line] (compute weights 3) |- (compute output 3);
    \path [line] (solve 3) -| (compute weights 3);
    \draw [line] (converged 3.east) -| node [near start] {\scriptsize yes}
                 ([xshift=2.25cm]done.east) -- (done.east);

\end{tikzpicture}
\caption{The staged training of the super-resolution RBFNN with the first stage in blue, the second stage in red and the last stage in green.}
\label{fig:flow_chart}
\end{figure}
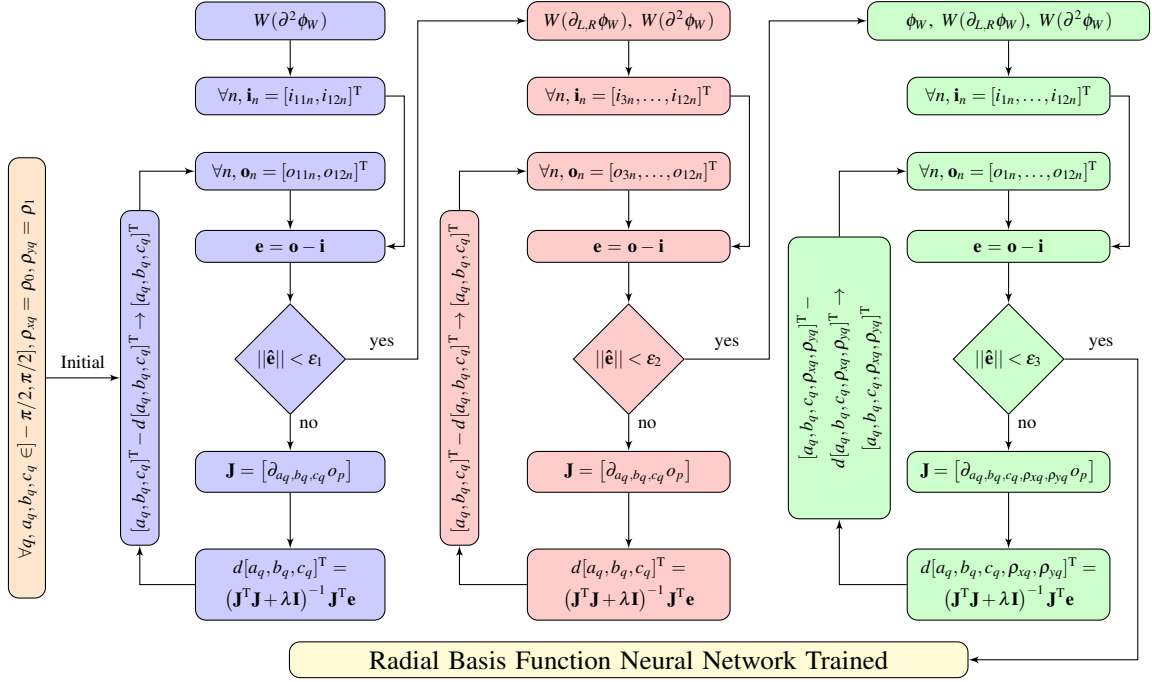

\subsection*{Multistage training}
\subsubsection*{The staged Levenberg-Marquardt algorithm}
The first step in the neural network training tries to match the network output to the second derivative of measured phase, using only the inputs $i_{11..12}$. Once the network is fully trained and the second derivatives of the output layer $\phi$ matches the second derivatives of $\phi_{GT}$ up to a small error, we restart the training process, but this time using the inputs $i_{3..12}$ to match the second derivative and the sine/cosine values of the first derivatives. We use the trigonometric functions to hide the discontinuities of the first derivatives of $\phi_W$ because trigonometric functions are differentiable and allow to compute the Jacobian matrix analytically. Once the network is trained (i.e. $e_{3..12}$ minimized for all neurons), the output layer should match the first central derivatives of $\phi_{GT}$ (as the training procedure matched both left and right derivatives, ultimately yielding the central derivative $\partial_C\phi_{GT}$) as well as the second derivatives. We finalize training the network using the all inputs $i_{1..12}$, including now the sine and cosine of $\phi_W$, to hide the phase jumps this time rather than their discontinuities inside their first derivatives. Once the errors $e_{1..12}$ are minimized, the output layer now matches the second and first derivatives of $\phi_{GT}$ as well as $\phi_{GT}$ itself.

The minimization procedure highlighted above will find the values of the basis function weights $u_n=(a_n,b_n,c_n,\rho_{x_n},\rho_{y_n})$ for all neurons $n\in\{1,..,N\}$ using a gradient-based algorithm. We used here the Levenberg-Marquardt Algorithm\cite{Levenb,Marq} (LMA), which minimizes the error $\mathbf{e}$ (not $\mathbf{\hat e}$) in the sense of the least square using the Jacobian matrix $\textbf{J}$. The solution is found by successive iterations, advancing the vector $\mathbf u=[u_1,\hdots,u_N]^\text{T}$ such that $\mathbf{u}^{new}=\mathbf{u}-\text{d}\mathbf{u}$  with 
$$\text{d}\mathbf u=\left(\mathbf{J^\text{T}J}+\lambda\mathbf{I}\right)^{-1}\mathbf{J^\text{T}e}.$$

The procedure to find $\lambda$ follows exactly the standard LMA and we only detail here the three training stages: 
\begin{enumerate}
    \item \textit{Matching the second derivatives of} $\phi_W$: The error vector is defined as $\mathbf e_n=[e_{11n},e_{12n}]^\text{T}=[o_{11n}-i_{11n},o_{12n}-i_{12n}]^\text{T}$. We only train the neural network to optimize the radial basis function weights $w_n$ at this stage using $\mathbf u_n=[u_{1n},u_{2n},u_{3n}]^\text{T}=[a_n,b_n,c_n]^\text{T}$. We have found that optimizing the activation distance early on does not really improve the quality of the output at this stage.  The quality of convergence at this stage is crucial to super-resolution. This stage is shown in blue in Fig. \ref{fig:flow_chart}.
    \item \textit{Matching the first and second derivatives of} $\phi_W$: The error vector is now redefined as $\mathbf e_n=[e_{3n},\hdots,e_{12n}]^\text{T}=[o_{3n}-i_{3n},\hdots,o_{12n}-i_{12n}]^\text{T}$. Here again, we train the neural network to optimize the radial basis function weights $w_n$ using $\mathbf u_n=[u_{1n},u_{2n},u_{n_3}]^\text{T}=[a_n,b_n,c_n]^\text{T}$. This stage propagates the super-resolution information to the first derivatives of the phase. This stage is shown in red in Fig. \ref{fig:flow_chart}.
    \item \textit{Matching $\phi_W$ as well as the first and second derivatives of $\phi_W$}: The error vector is defined as $\mathbf e_n=[e_{1n},\hdots,e_{12n}]^\text{T}=[o_{1n}-i_{1n},\hdots,o_{12n}-i_{12n}]^\text{T}$. We now optimize the neutral network to find the basis function weight $w_n$ and the inverse activation distances at this stage so $\mathbf u_n=[u_{1n},\hdots,u_{5n}]^\text{T}=[a_n,b_n,c_n,\rho_{x_n},\rho_{y_n}]^\text{T}$. This stage unwraps the phase globally, in one single sweep. This stage is shown in green in Fig. \ref{fig:flow_chart}.
\end{enumerate}

When super resolution is not needed, the training will only try to match the first left and right derivatives, together with the wrapped phase using only two training stages: 
\begin{enumerate}
    \item \textit{Matching the first derivatives of} $\phi_W$: The error vector is first define as $\ubar {\mathbf e}_n=[\ubar e_{3n},\hdots,\ubar e_{6n}]^\text{T}=[\ubar o_{3n}-\ubar i_{3n},\hdots,\ubar o_{6n}-\ubar i_{6n}]^\text{T}$. Here again, we train the neural network to optimize the radial basis function weights $w_n$ using $\mathbf u_n=[u_{1n},u_{2n},u_{n_3}]^\text{T}=[a_n,b_n,c_n]^\text{T}$. This stage propagates the super-resolution information to the first derivatives of the phase. This stage is shown in blue in Fig. \ref{fig:flow_chart_short}.
    \item \textit{Matching $\phi_W$ as well as the first derivatives of $\phi_W$}: The error vector is defined as $\ubar {\mathbf e}_n=[\ubar e_{1n},\hdots,\ubar e_{6n}]^\text{T}=[\ubar o_{1n}-\ubar i_{1n},\hdots,\ubar o_{6n}-\ubar i_{6n}]^\text{T}$. We now optimize the neutral network to find the actual basis function weight $w_n$ and activation distances at this stage so $\mathbf u_n=[u_{1n},\hdots,u_{5n}]^\text{T}=[a_n,b_n,c_n,\rho_{x_n},\rho_{y_n}]^\text{T}$. This stage unwraps the phase globally, in one single sweep. This stage is shown in red in Fig. \ref{fig:flow_chart_short}.
\end{enumerate}
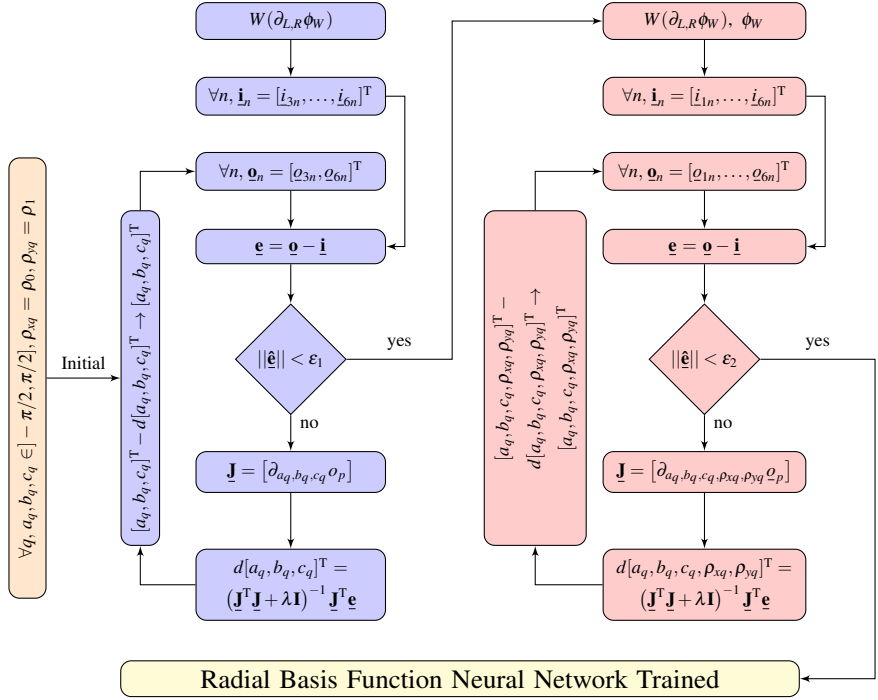
\begin{figure}[ht]
\centering
\begin{tikzpicture}[node distance=1cm,auto]
    \node [block 1] (stage 1 start) {\scriptsize $W(\partial_{L,R}\phi_W)$};
    \node [block 1, below of=stage 1 start] (compute input 1) {\scriptsize $\forall n,\,\ubar {\textbf{i}}_n=[\ubar i_{3n},\hdots,\ubar i_{6n}]^\text{T}$};
    \node [block 1, below of=compute input 1] (compute output 1) {\scriptsize $\forall n,\,\ubar {\textbf{o}}_n=[\ubar o_{3n},\ubar o_{6n}]^\text{T}$};
    \node [block 1, below of=compute output 1] (compute error 1) {\scriptsize $\ubar {\textbf{e}}=\ubar {\textbf{o}}-\ubar {\textbf{i}}$};
    \node [decision 1, below of=compute error 1] (converged 1) {\scriptsize $||\ubar {\mathbf{\hat e}}||<\epsilon_1$};
    \node [block 1, below of=converged 1,yshift=-.5 cm] (compute Jacobian matrix 1) {\scriptsize $\ubar{\textbf{J}}=\left[\partial_{a_q,b_q,c_q}o_p\right]$};
    \node [block 1, below of=compute Jacobian matrix 1,yshift=-0.5cm] (solve 1) {\scriptsize $d[a_q,b_q,c_q]^\text{T}=\left(\ubar{\textbf{J}}^\text{T}\ubar{\textbf{J}}+\lambda\textbf{I}\right)^{-1}\ubar{\textbf{J}}^\text{T}\ubar{\textbf{e}}$};
    \node [long block 1, left of=converged 1,xshift = -1 cm,yshift=-.25cm,rotate=90] (compute weights 1) {\scriptsize $ [a_q,b_q,c_q]^\text{T}-d[a_q,b_q,c_q]^\text{T}\rightarrow[a_q,b_q,c_q]^\text{T}$};
    \node [long block 0, left of=compute weights 1,xshift =- .5 cm,rotate=90] (initialize weights) {\scriptsize $ \forall q,\,a_q,b_q,c_q\in]-\pi/2,\pi/2],\,\rho_{xq}=\rho_0,\,\rho_{yq}=\rho_1$};
    \path [line] (stage 1 start) -- (compute input 1);
    \path [line] (initialize weights) -- node  {\scriptsize Initial} (compute weights 1);
    \draw [line] (compute input 1.east) -| ([xshift=0.25cm]compute input 1.east) |- (compute error 1.east);
    \path [line] (compute output 1) -- (compute error 1);
    \path [line] (compute error 1) -- (converged 1);
    \path [line] (compute Jacobian matrix 1) -- (solve 1);
    \path [line] (converged 1) -- node [near start] {\scriptsize no} (compute Jacobian matrix 1);
    \path [line] (compute weights 1) |- (compute output 1);
    \path [line] (solve 1) -| (compute weights 1);
    \node [block 2,right of =stage 1 start,xshift=4.5cm] (stage 2 start) {\scriptsize $W(\partial_{L,R}\phi_W)$, $\phi_W$};
    \node [block 2, below of= stage 2 start] (compute input 2) {\scriptsize $\forall n,\,\ubar{\textbf{i}}_n=[\ubar i_{1n},\hdots,\ubar i_{6n}]^\text{T}$};
    \node [block 2, below of=compute input 2] (compute output 2) {\scriptsize $\forall n,\,\ubar{\textbf{o}}_n=[\ubar o_{1n},\hdots,\ubar o_{6n}]^\text{T}$};
    \node [block 2, below of=compute output 2] (compute error 2) {\scriptsize $\ubar{\textbf{e}}=\ubar{\textbf{o}}-\ubar{\textbf{i}}$};
    \node [decision 2, below of=compute error 2] (converged 2) {\scriptsize $||\ubar{\mathbf{\hat e}}||<\epsilon_2$};
    \node [block 2, below of=converged 2,yshift=-.5 cm] (compute Jacobian matrix 2) {\scriptsize $\ubar{\textbf{J}}=\left[\partial_{a_q,b_q,c_q,\rho_{xq},\rho_{yq}}\ubar o_p\right]$};
    \node [block 2, below of=compute Jacobian matrix 2,yshift=-0.5cm] (solve 2) {\scriptsize $d[a_q,b_q,c_q,\rho_{xq},\rho_{yq}]^\text{T}=\left(\ubar{\textbf{J}}^\text{T}\ubar{\textbf{J}}+\lambda\textbf{I}\right)^{-1}\ubar{\textbf{J}}^\text{T}\ubar{\textbf{e}}$};
    \node [long block 2, left of=converged 2,xshift = -1.25 cm,yshift=-.25cm,rotate=90] (compute weights 2) {\scriptsize $ [a_q,b_q,c_q,\rho_{xq},\rho_{yq}]^\text{T}-d[a_q,b_q,c_q,\rho_{xq},\rho_{yq}]^\text{T}\rightarrow[a_q,b_q,c_q,\rho_{xq},\rho_{yq}]^\text{T}$};

    \draw [line] (converged 1.east) -| node [near start] {\scriptsize yes}
                 ([xshift=-2cm]stage 2 start.west) -- (stage 2 start.west);
    \path [line] (stage 2 start) -- (compute input 2);
    \draw [line] (compute input 2.east) -| ([xshift=0.25cm]compute input 2.east) |- (compute error 2.east);
    \path [line] (compute output 2) -- (compute error 2);
    \path [line] (compute error 2) -- (converged 2);
    \path [line] (compute Jacobian matrix 2) -- (solve 2);
    \path [line] (converged 2) -- node [near start] {\scriptsize no} (compute Jacobian matrix 2);
    \path [line] (compute weights 2) |- (compute output 2);
    \path [line] (solve 2) -| (compute weights 2);
    \node [block 4, below of=compute weights 2,xshift=-1cm,yshift=-3cm] (done) {Radial Basis Function Neural Network Trained};

    \draw [line] (converged 2.east) -| node [near start] {\scriptsize yes}
                 ([xshift=1cm]done.east) -- (done.east);

\end{tikzpicture}
\caption{Staged training of the RBFNN without super-resolution. The first stage is in blue and the second stage in red.}
\label{fig:flow_chart_short}
\end{figure}

\subsubsection*{Computation of the Jacobian matrix with super-resolution}
The Jacobian matrix used in the last stage of the training is given by
$$
    \mathbf{J}=\left[
    \begin{matrix}
        \partial_{u_1}\mathbf e_1&\hdots&\partial_{u_N}\mathbf e_1\\
        \vdots&\ddots&\vdots\\
        \partial_{u_1}\mathbf e_N&\hdots&\partial_{u_N}\mathbf e_N\\
    \end{matrix}
    \right]
\text{ where }
    \partial_{u_q} \mathbf e_p=\left[
    \begin{matrix}
        \partial_{a_q}o_{1p}&\partial_{b_q}o_{1p}&\partial_{c_q}o_{1p}&\partial_{\rho_{x_q}}o_{1p}&\partial_{\rho_{y_q}}o_{1p}\\
        \vdots&\vdots&\vdots&\vdots&\vdots\\
        \partial_{a_q}o_{12p}&\partial_{b_q}o_{12p}&\partial_{c_q}o_{12p}&\partial_{\rho_{x_q}}o_{12p}&\partial_{\rho_{y_q}}o_{12p}\\
    \end{matrix}
    \right].
$$

Here the matrix $\partial_{u_q} e_p$ corresponds to the partial derivative of the error $e_p$ between the output layer and the input layer computed at the $p^{th}$ neuron with respect to the weights $u_q$ of the $q^{th}$ neuron. Since the input layer does not depend on any neuron weights, the input values $i_{1p}$ to $i_{12p}$ have been dropped inside the partial derivatives and only the output values $o_{1p}$ to $o_{12p}$ were retained. To form the smaller Jacobian matrix matrices necessary to the first two stages, we just need to drop the corresponding terms in the full matrix $\partial_{u_q} e_p$, leading to
$$
    \text{First stage: }
    \partial_{u_q} \mathbf e_p=\left[
    \begin{matrix}
        \partial_{a_q}o_{11p}&\partial_{b_q}o_{11p}&\partial_{c_q}o_{11p}\\
        \partial_{a_q}o_{12p}&\partial_{b_q}o_{12p}&\partial_{c_q}o_{12p}\\
    \end{matrix}
    \right]
\text{ and Second stage: }
    \partial_{u_q}\mathbf e_p=\left[
    \begin{matrix}
        \partial_{a_q}o_{3p}&\partial_{b_q}o_{3p}&\partial_{c_q}o_{3p}\\
        \vdots&\vdots&\vdots\\
        \partial_{a_q}o_{12p}&\partial_{b_q}o_{12p}&\partial_{c_q}o_{12p}\\
    \end{matrix}
    \right].
$$

All the functions used in $o_{1_n}$ to $o_{12_n}$ are analytical and can be differentiated, since the wrapping operator $W$ was dropped from $o_{11_n}$ and $o_{12_n}$ using the condition $\partial^2\phi\in]-\pi,\pi]$.
We can now compute the Jacobian matrix elements taking the partial derivative on every term in Eq. \ref{eq:outcomp} with respect to $\omega\in\{a_q,b_q,c_q,\rho_{x_q},\rho_{y_q}\}$
\begin{equation}\label{eq:jac_der}
    \begin{matrix}
    &&\partial_{\omega }o_3=&-\partial_{x\omega }\phi\sin(\partial_x\phi)
    &\partial_{\omega }o_7=&-\partial_{x\omega }\phi\sin(\partial_x\phi)\\
    \partial_{\omega }o_1=&-\partial_{\omega }\phi\sin(\phi)
    &\partial_{\omega }o_4=&\partial_{x\omega }\phi\cos(\partial_x\phi)
    &\partial_{\omega }o_8=&\partial_{x\omega }\phi\cos(\partial_x\phi)
    &\partial_{\omega }o_{11}=&\partial_{xx\omega }\phi\\
    \partial_{\omega }o_2=&\partial_{\omega }\phi\cos(\phi)
    &\partial_{\omega }o_5=&-\partial_{y\omega }\phi\sin(\partial_y\phi)
    &\partial_{\omega }o_9=&-\partial_{y\omega }\phi\sin(\partial_y\phi)
    &\partial_{\omega }o_{12}=&\partial_{yy\omega }\phi\\
    &&\partial_{\omega }o_6=&\partial_{y\omega }\phi\cos(\partial_y\phi)
    &\partial_{\omega }o_{10}=&\partial_{y\omega }\phi\cos(\partial_y\phi)\\
    \end{matrix}
\end{equation}
The values of the partial derivatives used in Eq. \ref{eq:jac_der} are listed in the Methods section. 
\subsubsection*{Computation of the Jacobian matrix without super-resolution}
When dropping super-resolution, the Jacobian matrix $\ubar {\mathbf{J}}$ of Fig. \ref{fig:flow_chart_short} 
$$
    \ubar{\mathbf{J}}=\left[
    \begin{matrix}
        \partial_{u_1}\ubar {\mathbf e}_1&\hdots&\partial_{u_N}\ubar {\mathbf e}_1\\
        \vdots&\ddots&\vdots\\
        \partial_{u_1}\ubar {\mathbf e}_N&\hdots&\partial_{u_N}\ubar {\mathbf e}_N\\
    \end{matrix}
    \right]
$$
can be computed in a similar manner. Here the error $\partial_{u_q}\ubar {\mathbf e}_p$ is given by
$$
    \partial_{u_q} \ubar{\mathbf e}_p=\left[
    \begin{matrix}
        \partial_{a_q}\ubar o_{3p}&\partial_{b_q}\ubar o_{3p}&\partial_{c_q}\ubar o_{3p}\\
        \vdots&\vdots&\vdots\\
        \partial_{a_q}\ubar o_{6p}&\partial_{b_q}\ubar o_{6p}&\partial_{c_q}\ubar o_{6p}\\
    \end{matrix}
    \right]
$$
for the first stage and  
$$
    \partial_{u_q} \ubar{\mathbf e}_p=\left[
    \begin{matrix}
        \partial_{a_q}\ubar o_{1p}&\partial_{b_q}\ubar o_{1p}&\partial_{c_q}\ubar o_{1p}&\partial_{\rho_{x_q}}\ubar o_{1p}&\partial_{\rho_{y_q}}\ubar o_{1p}\\
        \vdots&\vdots&\vdots&\vdots&\vdots\\
        \partial_{a_q}\ubar o_{6p}&\partial_{b_q}\ubar o_{6p}&\partial_{c_q}\ubar o_{6p}&\partial_{\rho_{x_q}}\ubar o_{6p}&\partial_{\rho_{y_q}}\ubar o_{6p}\\
    \end{matrix}
    \right].
$$
for the second stage. 
\subsubsection*{Masking and clustering strategies}
With the main procedure highlighted, we can now focus on the initialization of our network, looking at masking, neuron clustering and receptor connections. The mask should be chosen before the training starts and should remain the same throughout the training. Most interferometry data carries noise, discontinuities, and regions that should be dropped from the interferogram. The mask should keep inside the input layer only the data that can be unwrapped with minimal error propagation. The mask over discarded data should slightly overlap with useful data. This strategy allows to compute properly phase derivatives at the mask boundary rather than using extrapolations. Further, the mask should neither split the data into separate regions nor have constricted regions.

The optimal number of receptors is integrated in the optimization procedure and does not have to be computed beforehand. Since we are using compact radial basis functions, any input $p$ such that $r_{pq}=[(x_p-x_q)^2\rho^2_{x_q}+(y_p-y_q)^2\rho^2_{y_q}]^{1/2}>1$ will not be connected to the neuron $q$. The training process is initialized by choosing arbitrary values for $\rho_{x_q}$ and $\rho_{y_q}$ and these values should be chosen carefully. In regions with rapid phase changes the activation distance inverses should be large. 

Neural networks often use a clustering method, such as k-means\cite{Forgy65,Lloyd82}, to improve the quality and speed of the training. However, the data pattern is rather inextricable \textit{a priori} without super-resolution, which is only gained \textit{a posteriori}. As a result, the shape of the mask and the distance between neurons, rather than the data inside the input layer, truly shapes neurons clustering in this work. This greatly simplifies the clustering procedure, which now boils down to a straightforward graph partitioning\cite{Karypis98} based on nearest-neighbor connections.

\subsection*{Parallel training}
Parallelization becomes necessary for moderately large dataset\cite{ribeiro2018,gunther2020}, as the size of the Jacobian matrix $\mathbf J$, even sparse, could be difficult to handle on today's supercomputers. This is especially true for high resolution two-dimensional interferograms obtained when measuring the electron density of high energy density plasmas. The basic clustering strategy described above can be used to split the main network into $K$ non-overlapping networks. As it is often the case with parallel codes, we introduce \textit{ghost} neurons\cite{long2008scalable}, which are duplicated neurons shared by exactly two networks. since the training of each network is now done independently, a synchronization step is require to make sure that all the output layer match seamlessly.

We used a single-nearest-neighbor search to define a single-layer of ghost neurons at the boundary between each clusters, allowing for some overlap between networks so that output layers can match seamlessly after synchronization. However, the synchronization procedure needs to keep very few of these ghost neurons to "stitch" the domains together.



\subsubsection*{The output layer}
The synchronization uses a constant phase $\Phi_k$, which is added to the output layer of the network  $k\in\{1,\dots,K\}$ as 
\begin{equation}\label{eq:full_output_layer}
  \phi_k(x,y)=\Phi_k+\sum_{q=M_k}^{N_{k}} w_q\psi \left(r_q\right)+\sum_{q=M_{gk}}^{N_{gk}} w_q\psi \left(r_q\right),  
\end{equation}
where the neurons $\{M_{k},\hdots,N_{k}\}$ are the neurons only owned by the $k^{th}$ network, while the neurons $\{M_{gk},\hdots,N_{gk}\}$ are the ghost neurons of the $k^{th}$ network, owned by the neighbors of the $k^{th}$ network. The last two terms of Eq. \ref{eq:full_output_layer} are the non-synchronized output of the RBFNN obtained using Eq. \ref{eq:rbf_form}. As the synchronization focuses solely on $\Phi_k$, the network parameters $a_k,\,b_k,\,c_k,\,\rho_{x_k}$, and $\rho_{y_k}$ are kept constant here and the last two terms of Eq. \ref{eq:full_output_layer} need to be computed once throughout the synchronization procedure.

\subsubsection*{The input layer}
For any ghost neuron $q$ shared with the network $k$ but owned by the network labelled $l_{qk}$, the value $\phi_k(x_q,y_q)$ might be initially different from the value $\phi_{l_{qk}}(x_q,y_q)$ when the staged training is over. Yet, we can synchronize the output layers across the different networks by simply defining the synchronization input layer of the $k^{th}$ network as
\begin{equation}\label{eq:par_input}
\begin{aligned}
    I_{qk}=&\phi_{l_{qk}}(x_q,y_q),
\end{aligned}
\end{equation}
with corresponding output value  
\begin{equation}
\begin{aligned}
    O_{qk}=&\phi_k(x_q,y_q).
\end{aligned}
\end{equation}
There is no need to use wrapping functions like sine or cosine here since we are dealing with a phase that has been unwrapped successfully for each separate networks but remains out of synchronization across the domain. Now, the error to minimize is given by
\begin{equation}
    \mathbf{E}^\text{T}\mathbf{E}=\sum_{k=1}^K\sum_{q=M_{gk}}^{N_{gk}}E^2_{qk}\text{ with }E_{qk}=O_{qk}-I_{qk}
\end{equation}
The error $E_{qk}$ is the squared difference of the input layer value from Eq. \ref{eq:par_input} computed at the location $(x_q,y_q)$ inside the $k^{th}$ cluster, i.e. $I_{qk}$, and the output layer value computed at the same location, i.e. $O_{qk}$.  

\subsubsection*{The synchronization Levenberg-Marquardt algorithm}

We use again the Levenberg-Marquardt algorithm to minimize the error $\mathbf{E}$ in the sense of the least square using the synchronization Jacobian matrix 
\begin{equation}
    \mathbf{J_s}=\left[
    \begin{matrix}
        \partial_{\Phi_1}E_{M_{g1}}&\hdots&\partial_{\Phi_K}E_{M_{g1}}\\
        \vdots&\ddots&\vdots\\
        \partial_{\Phi_1}E_{N_{gK}}&\hdots&\partial_{\Phi_K}E_{N_{gK}}\\
    \end{matrix}
    \right]
\end{equation}
Here we cannot drop the input values $I_{qk}$ from the Jacobian matrix since the input layer for the $k^{th}$ network may depend on a phase bias $\Phi_{k'}$ when ghost neurons in the network $k$ are owned by the network $k'$. We are now using a standard Levenberg-Marquardt algorithm to solve this problem. One final parallel third-stage training can be used after the synchronization procedure to eliminate any residual errors, while keeping all $\Phi_k$ constant. 


\subsection*{Accuracy of the staged neural network using synthetic phase}
This section presents the performance of the neural network for different types of synthetic phase variation with strong local aliasing. The first test looks at smoothly varying phase. Then, we focus on phase that varies randomly. The non-monotonic nature of the phase variation creates a new set of challenges on top of phase aliasing, especially in the presence of a fragmented mask and high noise levels. We looked at the accuracy of the neural network by computing the error between the ground-truth phase and the output layer, $\epsilon(x,y)=|\phi_{GT}(x,y)-\phi(x,y)|/(2\pi)$, which is given in units of $2\pi$ rather than radians and represents the normalized error with respect to the wrapped phase $\phi_W$, which spans an interval of $2\pi$. Each network is trained until the maximum error goes below $10^{-3}$ or when the overall error cannot be improved.  

\subsubsection*{Quasi-monotonic phase}
\begin{figure}[ht]
\centering
\includegraphics[width=4in]{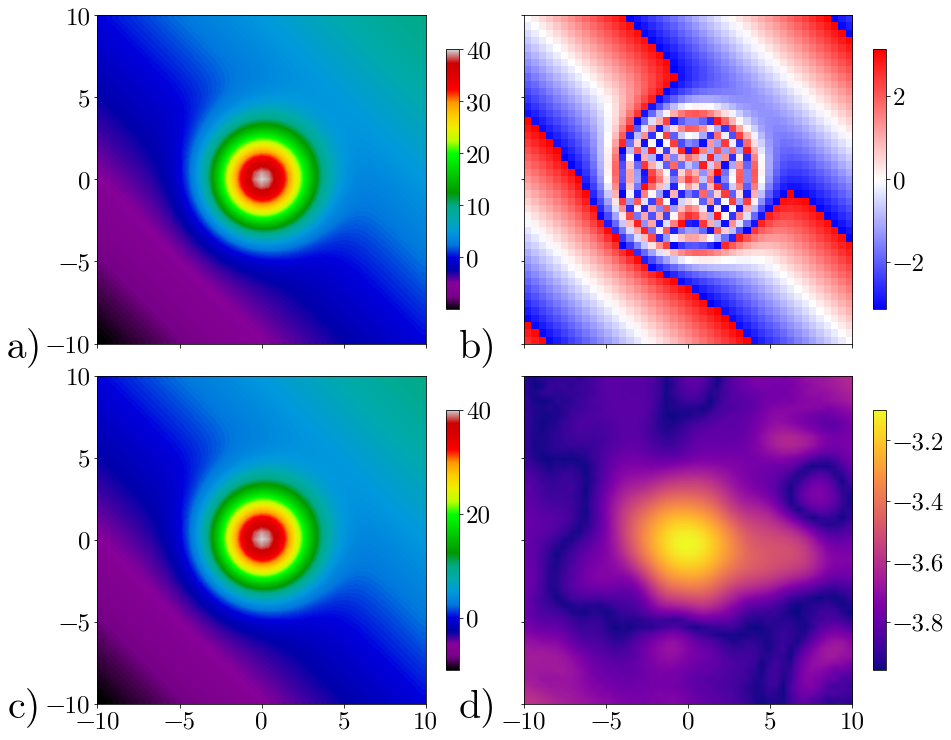}
\caption{a) The ground truth phase, b) the digitized wrapped phase, c) the RBFNN output, all in radians, and d) the output phase error on the $\log_{10}$ scale.}
\label{fig:smooth_phase_high}
\end{figure}

The quasi-monotonic phase is given by
\begin{equation}
    \phi_{GT}(x,y)=\alpha(x+y)+\beta\exp\left(-\frac{x^2+y^2}{\sigma^2}\right).
\end{equation}
Fig. \ref{fig:smooth_phase_high}-a shows the initial ground truth phase and Fig. \ref{fig:smooth_phase_high}-b the digitized wrapped phase with strong aliasing, all in radians. The neural network output layer is virtually identical to $\phi_{GT}$. However, the very high accuracy is obtained only after removing a constant bias that exist between the two phases. This bias is not an error. Rather it comes from a lack of absolute reference between the two phases. Since this bias cannot be determined from the wrapped phase shown in Fig. \ref{fig:smooth_phase_high}-b, we computed this bias to make the average of network output equal to the average of the ground truth and the recovered phase is shown in Fig. \ref{fig:smooth_phase_high}-c. In reality, we would not have access to this information when performing real phase measurements. But this limitation is physical rather than imposed by the method presented here. For $\beta<40$, the RBFNN recovers the ground-truth phase from the digitized phase with an error well below $10^{-3}$. The error becomes quickly worse with larger values of $\beta$. After this correction, Fig. \ref{fig:smooth_phase_high}-d shows that the maximum error between the RBFNN and the initial phase is less than 0.1\%. 

\subsubsection*{Random phase with masked data}

\begin{figure}[ht]
\centering
\includegraphics[width=4.5in]{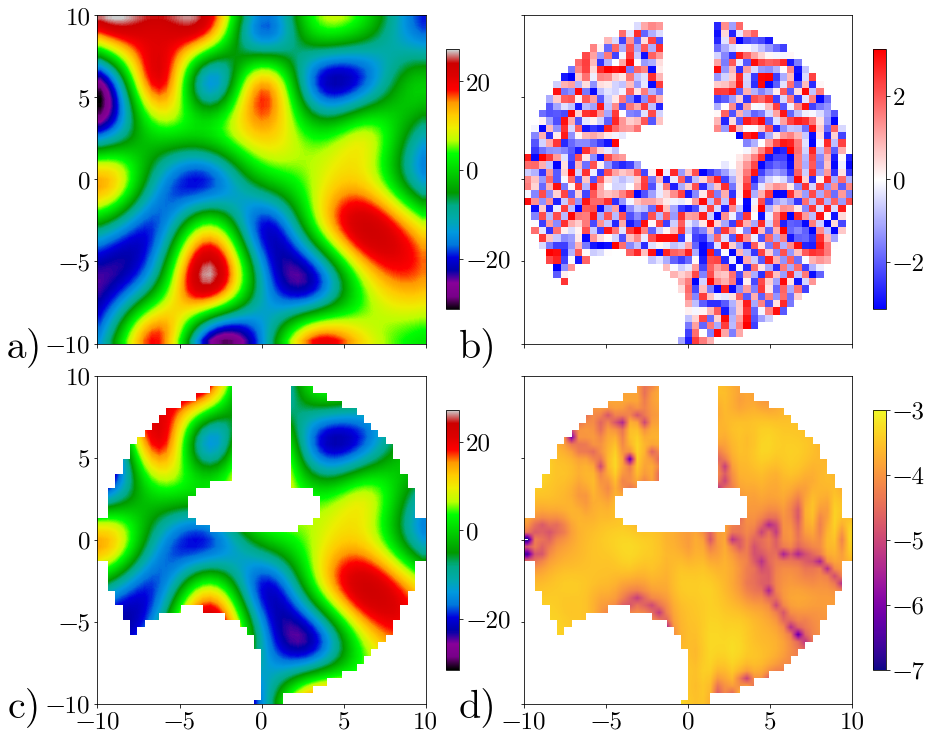}
\caption{a) The ground truth phase, neural network output phase and wrapped phase of Fig. \ref{fig:smooth_phase_high} along the \textit{x}-direction, together with their b) first and c) second derivatives along the \textit{x}-direction with mask. d) The output phase error on the $\log_{10}$ scale.}
\label{fig:random_phase_high_masked}
\end{figure}

\begin{figure}[ht]
\centering
\includegraphics[width=2.75in]{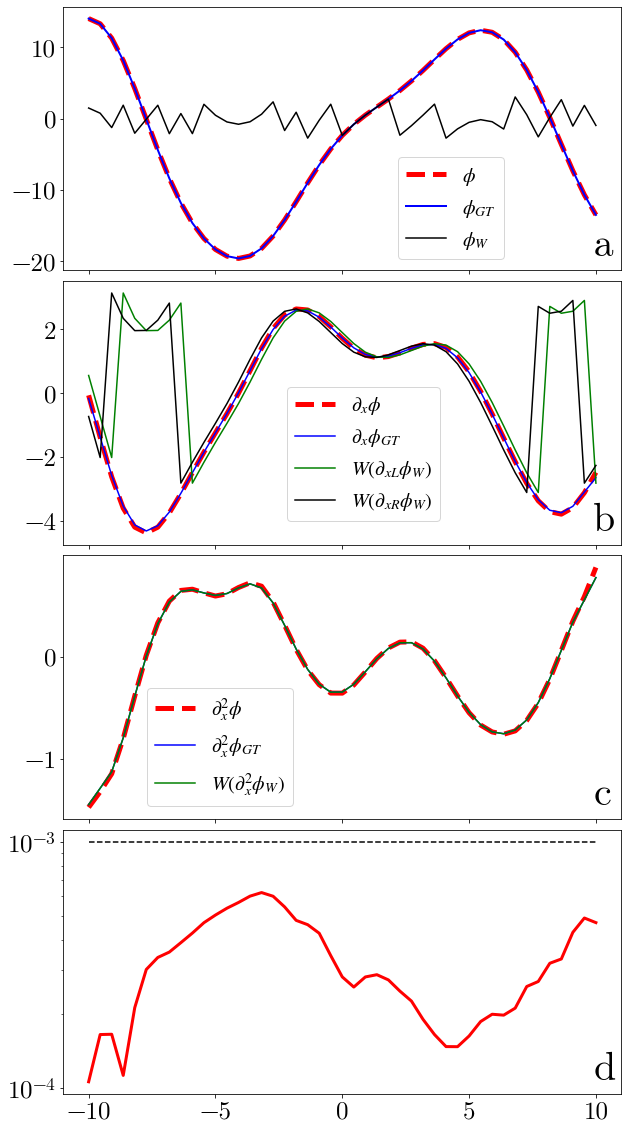}
\caption{a) The ground truth phase, neural network output phase and wrapped phase of Fig. \ref{fig:smooth_phase_high} along the \textit{x}-direction, together with their b) first and c) second derivatives along the \textit{x}-direction. d) The output phase error on the $\log_{10}$ scale.}
\label{fig:random_phase_high_1D}
\end{figure}

When the phase varies randomly across the domain, the neural network cannot exploit any trend to recover the ground truth $\phi_{GT}$.  If aliasing is introduced, then it becomes very difficult to even attempt the task manually.  While Fig. \ref{fig:random_phase_high_masked}-a shows that $\phi_{GT}$ does not vary wildly, the digitized, wrapped phase in Fig. \ref{fig:random_phase_high_masked}-b shows that a randomly varying phase is in fact relatively difficult to unwrap. Yet the output of the neural network shown in Fig. \ref{fig:random_phase_high_masked}-c matches well $\phi_{GT}$, with and error below 0.1\% shown in Fig. \ref{fig:random_phase_high_masked}-d. The error is more homogeneously distributed compared to the quasi monotonic phase presented in the previous section, mostly caused by global (rather than local) aliasing. There is very little change of the overall error compared to the unmasked case (not shown). Fig. \ref{fig:random_phase_high_1D}-a shows that aliasing is large enough to cause the wrapped phase to increase smoothly, while the ground-truth phase actually decreases. This happens in regions where the first derivative of $\phi_{GT}$, shown in Fig. \ref{fig:random_phase_high_1D}-b, is smaller than $-\pi$, causing $W\left(\partial\phi_W\right)$ to wrap around. Note that this wrapping is not problematic since we are using the sine and cosine functions when training our neural network on first derivatives, which continuously vary throughout phase jumps. 

Since the neural network is trained on a dataset that contains the first and second derivatives of the phase, we can take the derivatives of the neural network output to estimate the derivatives of the phase. Fig. \ref{fig:random_phase_high_1D}-b shows an excellent agreement with the ground truth phase derivative. We clearly see here that the RBFNN cannot match the left and right first derivatives simultaneously, since they have different values. Rather the RBFNN matches the average, which is the central first derivative. As shown in Eq. \ref{eq:inlayer}, the neural network uses the left and right derivatives of $\phi_W$ to compute the weights use in the output layer. So, the derivative of $\phi$, which also matches the derivative of $\phi_{GT}$, is located in between the left and right derivatives of $\phi_W$, as expected (see Fig. \ref{fig:random_phase_high_1D}-b). As a result, using the error $\mathbf{\hat e}$ given in Eq. \ref{eq:convergence_error} makes more sense than using $\mathbf{e}$. Based on the assumptions that $\phi_{GT}\in]-\pi,\pi]$ and $\phi_{GT}$ is continuous, we see that $W(\partial^2_x\phi_{W})$ has no jump since $W(\partial^2_x\phi_{W})=\partial^2_x\phi_{GT}$. 

Fig. \ref{fig:random_phase_high_1D} shows clearly how the neural network can recover the ground truth $\phi_{GT}$, without explicitly unwrapping it. The output of the neural network and its derivatives are continuous by construction, since they are the sum of continuous radial basis functions. At the end of the first stage of the training, the second derivative of the neural network matches directly the second derivative of the wrapped phase, which is continuous since $W(\partial^2_x\phi_{W})=\partial^2_x\phi_{GT}$ and $\partial^2_x\phi_{GT}$ is continuous.  At the end of the first stage, the output of the neural network is continuous since the output is continuous by construction. At the end of second stage, the network output matches the first derivatives of the wrapped phase via the sine and cosine functions. This approach hides the phase jumps created by the wrapping operator $W$ when aliasing exists.  Again, at the end of this stage, the output of the neural network is also continuous since it is the sum of continuous functions.  During the third stage, where the network is trained to match the wrapped phase values via the sine and cosine functions, its output again remains continuous. So, the training process forced the output of the neural network to match the sine and cosine of the wrapped phase and the radial basis functions used to build this network forced the output to be continuous, allowing to remove the jumps of the wrapped phase.
\begin{figure}[ht]
\centering
\includegraphics[width=4in]{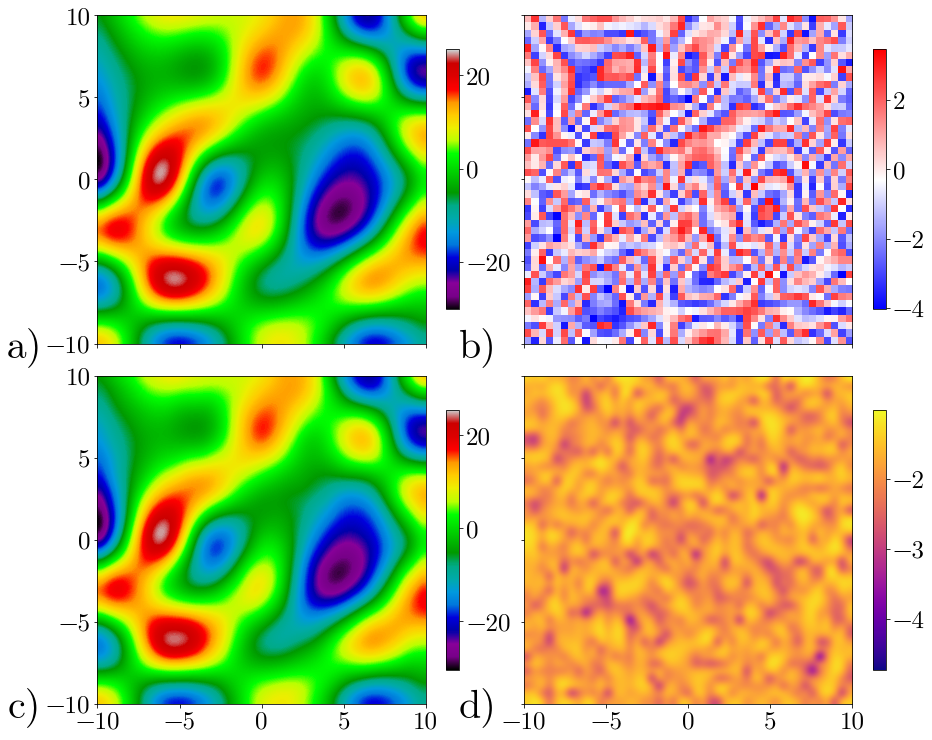}
\caption{a) The ground truth phase, b) the digitized wrapped phase, c) the RBFNN output, all in radians, and d) the phase error on the $\log_{10}$ scale with a noise level $\gamma=0.1$ or 10\% of the maximum value of the wrapped phase.}
\label{fig:random_phase_noise_high}
\end{figure}

\begin{figure}[ht]
\centering
\includegraphics[width=2.75in]{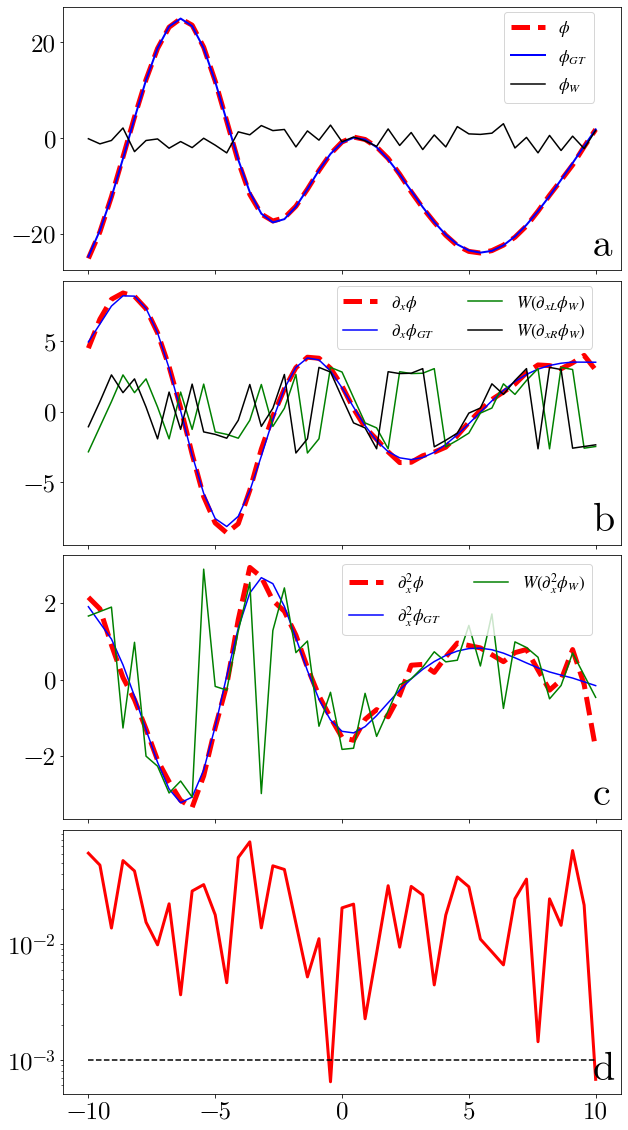}
\caption{a) The ground truth phase, neural network output phase and wrapped phase of Fig. \ref{fig:random_phase_noise_high} along the \textit{x}-direction with a noise level $\gamma=0.1$ or 10\% of the maximum value of the wrapped phase. Their b) first and c) second derivatives along the \textit{x}-direction. d) The output phase error on the $\log_{10}$ scale.}
\label{fig:random_phase_noise_high_1D}
\end{figure}

\subsubsection*{Random phase with noise}
When there is no aliasing, the noise can be removed from the wrapped phase using standard filtering techniques specifically developed for interferograms, such as the fringe smoothing approach\cite{berardino2002}, local fringe frequency estimation\cite{feng2016}, windowed Fourier filtering\cite{kemao2007,kemao2008}, or Gabor filter local frequency\cite{estrada2017}. Any of these techniques can be applied to the wrapped phase before feeding it to the neural network. When filtering the wrapped phase, we can detect locations with noisy data by computing the phase residues and mask out locations where the residues leads to a non conservative result\cite{goldstein1988}, providing that the ground truth phase is conservative (e.g. interferogram of topographic data).  Filtering can also be done during the unwrapping procedure \cite{flynn1997,costantini1998,fornaro2005,yu2011,tayebi2020} but cannot be applied here as the filtering procedure is deeply dependant of the unwrapping method. However, when aliasing is present, direct filtering becomes more problematic. For one, the method of residue cannot be used reliably. Furthermore, aliasing can behave like noise and it becomes difficult to differentiate between good data that was wrapped multiple times and noisy data. 

To look at the impact of noise on the neural network performance for strongly aliased phase, we added noise $\mathcal{N}(x,y)\in[-1,1]$ to the wrapped phase as
\begin{equation}
    \phi_W(x,y)=W\left(\phi_{GT}(x,y)\right)+\gamma\pi\mathcal{N}(x,y)
\end{equation}
where $\gamma$ is a constant controlling the maximum noise level. Fig. \ref{fig:random_phase_noise_high} shows that the neural network recovers the ground truth $\phi_{GT}$ with an error that is on the order of the noise level added to the wrapped phase. The neural network tends to perform well for $\gamma<0.1$, but tend to develop O(1) error when $\gamma>0.1$. Similar results are found with masked data. So, without specific noise filtering strategy working on the aliased wrapped phase, we find that the neural network remains reliable for noise levels below 10\% of the wrapped phase. Larger noise levels will require some filtering beforehand. Fig. \ref{fig:random_phase_noise_high_1D} shows how the regularization avoids over-fitting of the network output, limiting the impact of the noise on unwrapped phase.

\section*{Accuracy of the staged neural network training using interferometry data}
 After a series of test used to determine the accuracy of the RBFNN and presented in the \textit{Methods} section, we now use the proposed staged training on real interferograms generated by the interference of a green laser beam with a high energy density plasma\cite{drake2018}.  The phase shift corresponds to the line-average electron density\cite{hutchinson2002} of the plasma.  The plasmas were generated by using a multipin radial foil configuration\cite{radial_foils2014} connected to the electrodes of a pulsed-power driver\cite{greenly2008}. In this case, we do not know the ground-truth and we assessed the quality of the unwrapping procedure by looking at the difference between the measured wrapped phase and the neural network output layer. The final error, $\epsilon(x,y)=|W(\phi_W(x,y))-W(\phi(x,y)|)/(2\pi)$, is given in units of $2\pi$. It is the normalized error with respect to the wrapped phase $\phi_W$, which spans an interval of $2\pi$.


\begin{figure}[ht]
\centering
\includegraphics[width=6.75in]{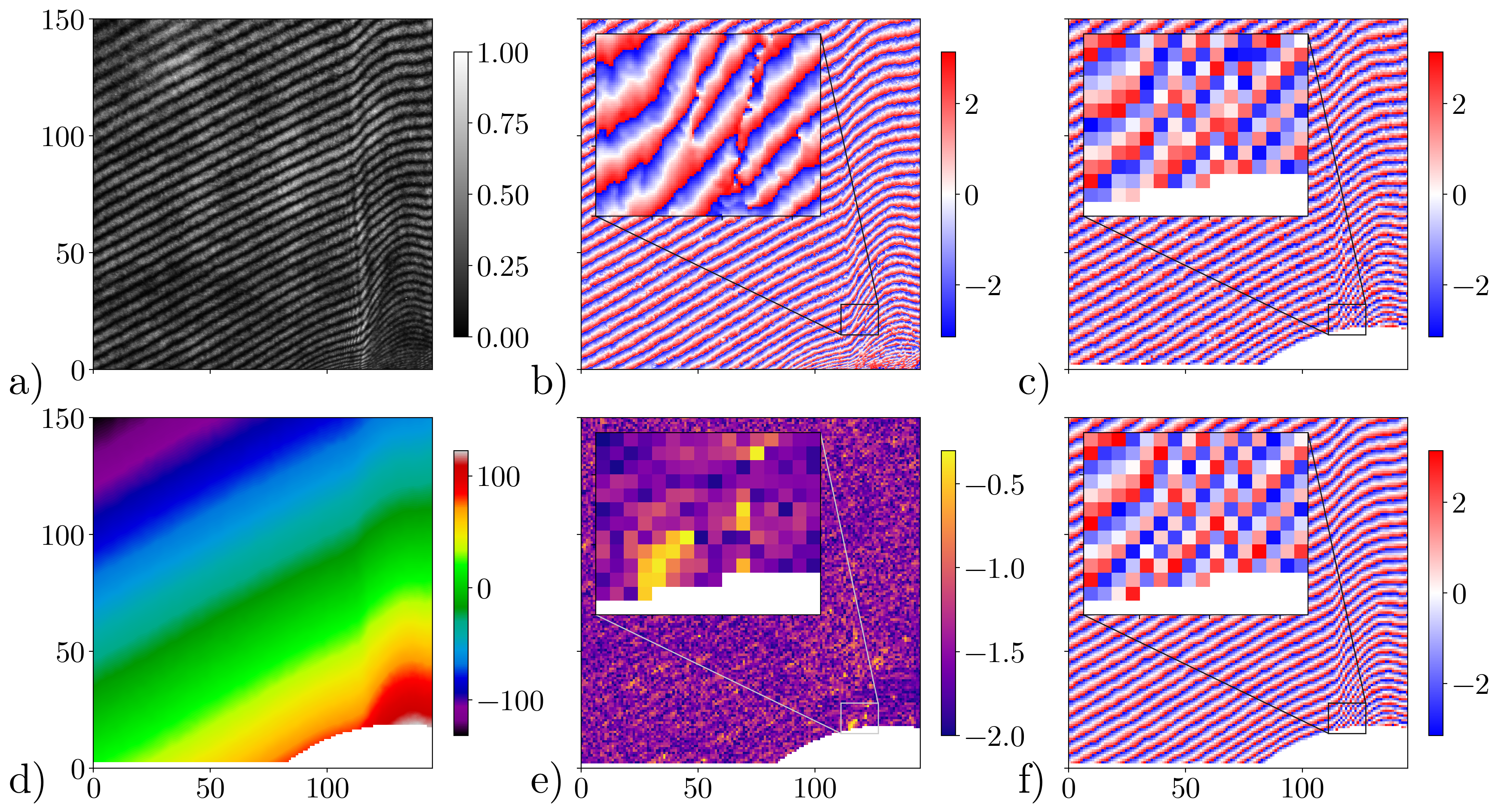}
\caption{a) Normalized interferometer of the left side of a hollow plasma jet. The jet is symmetric with respect to the right axis. b) The wrapped phase after applying Fourier filtering to keep the dominant modes. c) Digitized down-sampled phase with mask to hide the regions where phase data should not be used. d) The actual output $\phi$ of the RBFNN and e) the $\log_{10}$ error between the wrapped phase $\phi_W$ and $W(\phi)$. f) The wrapped RBFNN output is given for reference. All phase data are in radians. The zoomed panels highlight were the region with strongest aliasing.}
\label{fig:interferometer_data}
\end{figure}

\begin{figure}[ht]
\centering
\includegraphics[width=6.75in]{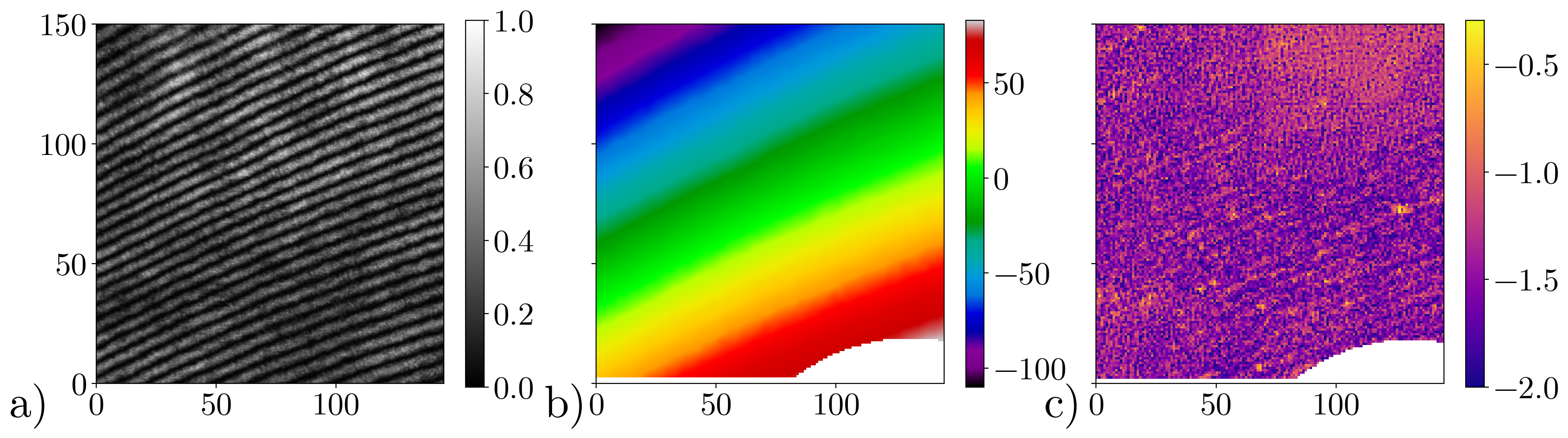}
\caption{a) Normalized interferometer with no plasma present, b) the RBFNN output in radians $\phi$ and c) the $\log_{10}$ error between the wrapped phase $\phi_W$ and $W(\phi)$.}
\label{fig:interferometer_data_bg}
\end{figure}

The interferogram is presented in Fig. \ref{fig:interferometer_data}-a. The measurement is based on shearing\cite{Sarkisov96} rather than Mach-Zehnder interferometry. The former uses a single reference path which is insensitive to mechanical vibrations, which greatly affects the fringe pattern of the latter.  As a result, it is possible to use a reference phase, by using phase data without plasma, and subtract it from the measurement done when a plasma is present.  The difference in phase is proportional to the line-average electron density. Starting with the region of interest shown in Fig. \ref{fig:interferometer_data}-a, the Fourier transform gives a spectrum that is symmetric with respect to the origin since the phase data is real valued. We use a single square filter to isolate the dominant modes, but excluding the origin, where the DC component is located. The inverse Fourier transform is now complex valued since the filter broke the symmetry with respect to the origin. The phase of each complex values corresponds to the wrapped phase measured by the interferogram\cite{Takeda:82,Macy:83,Roddier:87} and seen in Fig. \ref{fig:interferometer_data}-b. The data is then down-sampled by a factor of $6\times6$ to compress the interferogram (seen in Fig. \ref{fig:interferometer_data}-c). While the compression is not necessary to demonstrate the efficacy of the RBFNN, this compression created a region with strong aliasing (the zoomed portion of Fig. \ref{fig:interferometer_data}-c). A mask was used to drop the data where fringes could not be resolved clearly. We then trained the neural network with super-resolution. The output of the network is presented in Fig. \ref{fig:interferometer_data}-d. The bump in electron density caused the plasma jet appears clearly in the figure. The error between the measured wrapped phase of Fig. \ref{fig:interferometer_data}-c and the wrapped value of the output of the RBFNN of Fig. \ref{fig:interferometer_data}-d is on Fig.\ref{fig:interferometer_data}-e is on the order of 10\%, leading to an average error that is comparable to the noise recorded by the interferometer and clearly visible in the insert of Fig. \ref{fig:interferometer_data}-b. We see two types of error larger than 10 \% in this figure. The error that is randomly distributed throughout is caused by a local phase jump caused by the noise when the wrapped phase is close to to $-\pi$ or $\pi$, a noise that is not present in the output of the network due to regularization. The second type of error is closer to a true error, as the RBFNN has some difficulty to unwrap the phase accurately (region shown in the zoomed insert of Fig. \ref{fig:interferometer_data}-e). This error is coming from the $6\times6$ compression ratio, which has aliased the phase slightly beyond the capabilities of the RBFNN. However, this error can disappear if we use a $5\times5$ compression ratio. It is important to note that this error did not propagate to the neighboring neurons. If we consider the low, average error level of Fig. \ref{fig:interferometer_data}-e and the smoothness of the output, the RBFNN unwrapped the phase successfully. The wrapped output is shown in \ref{fig:interferometer_data}-f and can be compared to the measured phase in \ref{fig:interferometer_data}-c. Without super-resolution, the RBFNN was not able to unwrap the phase. 

\begin{figure}[ht]
\centering
\includegraphics[width=6.75in]{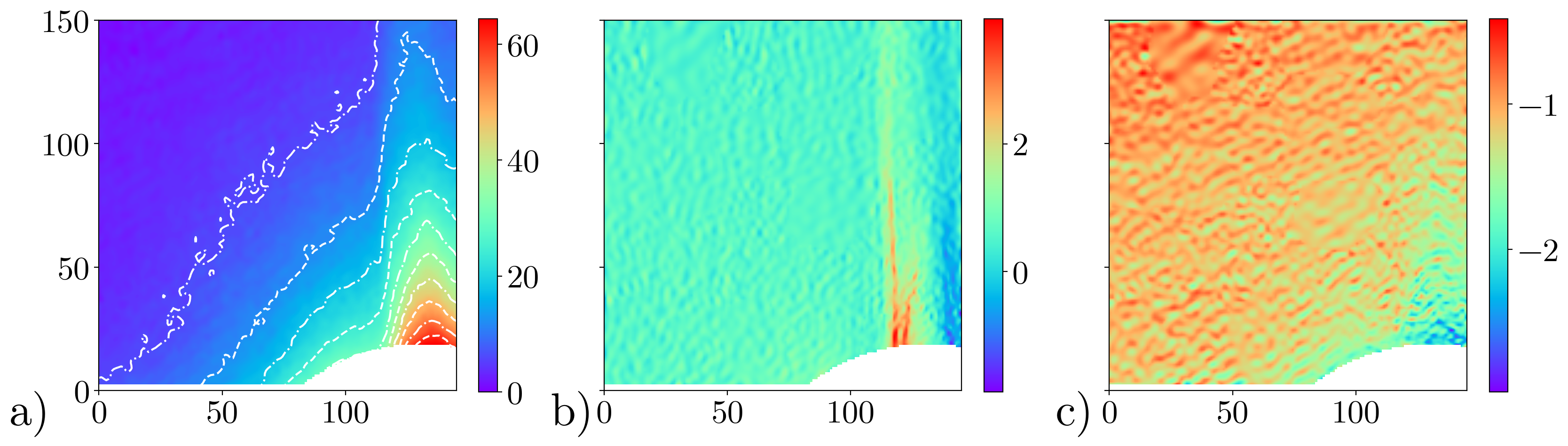}
\caption{a) The line average density of the hollow plasma jet (given in radian) and the derivatives along b) the horizontal and c) vertical directions in arbitrary units.  The axis of symmetry in to the right of each panel}
\label{fig:line_average}
\end{figure}

Since the shearing interferometer is mechanically stable, we can measure accurately the density of the jet by subtracting the background phase from Fig. \ref{fig:interferometer_data}-d. Following the exact same procedure, we can process the same region of the interferogram without any plasma. In this case, the fringe pattern is relatively periodic, as shown in Fig. \ref{fig:interferometer_data_bg}-a.  We get the wrapped background phase using the same Fourier filter as the one used for the interferogram with plasma. Since the pattern of the interferogram is clearly resolved, we trained the RBFNN without super-resolution, leading to the output presented in Fig. \ref{fig:interferometer_data_bg}-b. It is interesting to note that the error between the wrapped output layer and the data, shown in Fig. \ref{fig:interferometer_data_bg}-c, is similar to the error when the plasma is present. This indicates that the error is mostly caused by noise. Once the background phase is removed from the phase with plasma, we get the line average density of the jet presented in Fig. \ref{fig:line_average}-a. While noise is present in the density measurement, its source has been filtered by our earlier Fourier transform. We believe that the density fluctuations seen in the RBFNN output derivatives shown in Fig. \ref{fig:line_average}-b and c do not carry any physical information of the density itself. As a result, an Abel inversion technique that is robust to significant noise levels (e.g. Ref. \citenum{li2006}) should be used to compute the volume electron density. We can note the difference in smoothness between domain due to the optimization of the activation distance during the last stage.

\section*{Discussion}
The proposed RBFNN incorporates the functions necessary to deal with aliased interferograms by combining: 1-scattered neuron placement, allowing to discard relatively easily corrupted data while keeping data carrying high fidelity information; 2- the use of a mask to hide external geometries, which are often present in phase measurement; 3-a regularization scheme which can filter noise very effectively. The RBFNN can unwrapped the phase extracted from an interferogram by comparing measurements to the output of the RBFNN through sine and cosine functions. These functions hide the existence of any discontinuity in the wrapped phase from the training set. Taking into account that the RBFNN output is continuous by construction, the neural network yields a phase that is fully unwrapped once the error between the input and output layers has been minimized. As the network is trained to match the first and second derivatives of the phase, high-fidelity gradients can be computed directly from the RBFNN output since the impact of noise was limited by regularization. The network structure allow a clustering strategy where parallelization is easy to implement. It transforms a dense matrix into a block diagonal matrix, speeding up the training substantially. 

This work did not attempt to do any filtering in the pre-learning stage, except from a Fourier filter,which was mostly used to get the complex amplitude field, allowing to compute the wrapped phase readily. However, filtering techniques can be used in conjunction with the proposed algorithm. While regularization does filter data by limiting over-fitting, it should not be considered a very effective filter. First, the regularization parameter is global. Second, the regularization is static and there is no mechanism in place in our training that can optimize it. 

While the RBFNN presented here requires more memory and computational power than more basic phase unwrapping algorithm (e.g. Ref. \citenum{Ghiglia:94}), errors are relatively easy to detect and remain local, as shown in Fig. \ref{fig:interferometer_data}-e. Combined with the proposed parallelization strategy, the unwrapping time can be reduced substantially. While the training procedures with and without super-resolution are clearly separated in this work, it is possible to use phase derivative averages to find region where super-resolution is required (i.e. $|W(\phi_w)|>\pi$) and regions where its not (i.e. $|W(\phi_w)|<\pi$). The Jacobian matrix can be adapted locally to each method seamlessly. However, this criterion are not absolute and super-resolution should be used as much as possible. It is also possible to extend the method to three-dimensional interferograms easily, as discussed in the \textit{Method} section. At this point, a heavy use of block diagonalization is required to generate Jacobian matrices sparse enough to allow for reasonable training times. 
\section*{Data availability}
The datasets generated and analysed during the current study are not publicly available due to the continued evolution of the RBFNN but are available from the corresponding author upon request.
\section*{Methods}
\subsection*{Computation of the partial derivatives used inside the Jacobian matrices}
This section lists the analytic functions used to compute the Jacobian matrices used in this paper. The output layer is $\phi(x,y)=\sum_{q=1}^N w_q\psi \left(r_q\right) $ with $ r_q=\sqrt{(x-x_q)^2\rho^2_{x_q}+(y-y_q)^2\rho^2_{y_q}+\epsilon}$. Further, $\psi'=\frac{d}{dr}\psi$, $\psi''=\frac{d^2}{dr^2}\psi$ and $\psi'''=\frac{d^3}{dr^3}\psi$. $\epsilon$ is used in computations to avoid a possible division by zero, which only happens numerically. The problematic terms $w_q/r_q$ found below are multiplied by $\psi'$ or $\psi'''$, while $\lim_{r_q\to 0}\psi'\propto r_q$ and $\lim_{r_q\to 0}\psi'''\propto r_q$ for radial basis functions.
\subsubsection*{Partial derivatives with respect to the RBFNN parameters}
\begin{equation*}
    \partial_{a_q}\phi=\psi \left(r_q\right)
\end{equation*}
\begin{equation*}
    \partial_{b_q}\phi=\left(x-x_q\right) \psi \left(r_q\right)
\end{equation*}
\begin{equation*}
    \partial_{c_q}\phi=\left(y-y_q\right) \psi \left(r_q\right)
\end{equation*}
\begin{equation*}
    \partial_{\rho_{x_q}}\phi=\frac{w_q \left(x-x_q\right)^2 \rho_{x_q} }{r_q} \psi'\left(r_q\right)
\end{equation*}
\begin{equation*}
    \partial_{\rho_{y_q}}\phi=\frac{w_q \left(y-y_q\right)^2 \rho_{y_q}}{r_q} \psi'\left(r_q\right)
\end{equation*}

\subsubsection*{Partial derivatives along \textit{x}}
\begin{equation*}
    \partial_{x}\phi=\sum_{q=1}^N b_q \psi \left(r_q\right)+\frac{w_q \left(x-x_q\right) \rho _{x_q}^2 \psi '\left(r_q\right)}{r_q}
\end{equation*}
\begin{equation*}
    \partial_{xx}\phi=\sum_{q=1}^N\left[\frac{ \left(w_q+b_q \left(x-x_q\right)\right) \rho _{x_q}^2 }{r_q}-\frac{w_q \left(x-x_q\right){}^2 \rho _{x_q}^4 }{r_q^3}\right]\psi '\left(r_q\right)+\frac{w_q \left(x-x_q\right){}^2 \rho _{x_q}^4 }{r_q^2}\psi ''\left(r_q\right)
\end{equation*}
\begin{equation*}\partial_{xa_q}\phi=
\frac{ \left(x-x_q\right) \rho _{x_q}^2 }{r_q}\psi '\left(r_q\right)
\end{equation*}
\begin{equation*}\partial_{xxa_q}\phi=
\left[
\frac{\rho _{x_q}^2 }{r_q}
-\frac{\left(x-x_q\right)^2 \rho _{x_q}^4 }{r_q^3}
\right]\psi '\left(r_q\right)
+\frac{\left(x-x_q\right)^2 \rho _{x_q}^4 }{r_q^2}\psi ''\left(r_q\right)
\end{equation*}
\begin{equation*}\partial_{xb_q}\phi=
\psi \left(r_q\right)
+\frac{ \left(x-x_q\right)^2 \rho _{x_q}^2 }{r_q}\psi '\left(r_q\right)
\end{equation*}
\begin{equation*}\partial_{xxb_q}\phi=
\left[
 \frac{3 \left(x-x_q\right) \rho _{x_q}^2 }{r_q}
-\frac{\left(x-x_q\right)^3 \rho _{x_q}^4 }{r_q^3}
\right]\psi '\left(r_q\right)
+\frac{\left(x-x_q\right)^3 \rho _{x_q}^4 }{r_q^2}\psi ''\left(r_q\right)
\end{equation*}

\begin{equation*}\partial_{xc_q}\phi=
\frac{\left(x-x_q\right) \left(y-y_q\right) \rho _{x_q}^2 }{r_q}\psi '\left(r_q\right)
\end{equation*}
\begin{equation*}\partial_{xxc_q}\phi=
\left[
 \frac{\left(y-y_q\right) \rho _{x_q}^2 }{r_q}
-\frac{\left(x-x_q\right)^2 \left(y-y_q\right) \rho _{x_q}^4 }{r_q^3}
\right]\psi '\left(r_q\right)
+\frac{\left(x-x_q\right)^2 \left(y-y_q\right) \rho _{x_q}^4 }{r_q^2}\psi ''\left(r_q\right)
\end{equation*}

\begin{equation*}\partial_{x\rho_{x_q}}\phi=
\left[
 \frac{\left(2w_q+b_q\left(x-x_q\right)\right) \left(x-x_q\right) \rho _{x_q} }{r_q}
-\frac{w_q \left(x-x_q\right)^3 \rho _{x_q}^3 }{r_q^3}
\right]\psi '\left(r_q\right)
+\frac{w_q \left(x-x_q\right)^3 \rho _{x_q}^3 }{r_q^2}\psi ''\left(r_q\right)
\end{equation*}
\begin{equation*}
\begin{aligned}
\partial_{xx\rho_{x_q}}\phi=
&-\left[
 -\frac{2\left(w_q+2 b_q \left(x-x_q\right)\right) \rho _{x_q} }{r_q}
+\frac{\left(5 w_q + 2 b_q \left(x-x_q\right)\right)\left(x-x_q\right)^2 \rho_{x_q}^3 }{r_q^3}
-\frac{3 w_q \left(x-x_q\right)^4 \rho _{x_q}^5 }{r_q^5}
\right]\psi '\left(r_q\right)\\
&+\left[
\frac{\left(5 w_q + 2 b_q \left(x-x_q\right)\right)\left(x-x_q\right)^2  \rho _{x_q}^3 }{r_q^2}
-\frac{3 w_q \left(x-x_q\right)^4 \rho _{x_q}^5 }{r_q^4}
\right]\psi ''\left(r_q\right)\\
&+\frac{ w_q \left(x-x_q\right)^4  \rho _{x_q}^5}{r_q^3}\psi '''\left(r_q\right)
\end{aligned}
\end{equation*}
\begin{equation*}
    \partial_{x\rho_{y_q}}\phi=
    \left[\frac{b_q \left(y-y_q\right)^2 \rho _{y_q} }{r_q}-\frac{w_q \left(x-x_q\right) \left(y-y_q\right)^2 \rho _{x_q}^2 \rho _{y_q} }{r_q^3}\right]\psi '\left(r_q\right)+\frac{w_q \left(x-x_q\right) \left(y-y_q\right)^2 \rho _{x_q}^2 \rho _{y_q} }{r_q^2}\psi ''\left(r_q\right)
\end{equation*}
\begin{equation*}
\begin{aligned}
    \partial_{xx\rho_{y_q}}\phi=
    &-\left[\frac{\left(w_q+2 b_q \left(x-x_q\right) \right)\left(y-y_q\right)^2 \rho _{x_q}^2 \rho _{y_q}}{r_q^3}-\frac{3 w_q \left(x-x_q\right)^2 \left(y-y_q\right)^2 \rho _{x_q}^4 \rho _{y_q} }{r_q^5}\right]\psi '\left(r_q\right)\\
    &+\left[\frac{\left(w_q+2 b_q \left(x-x_q\right) \right)\left(y-y_q\right)^2 \rho _{x_q}^2 \rho _{y_q}}{r_q^2}-\frac{3 w_q \left(x-x_q\right)^2 \left(y-y_q\right)^2 \rho _{x_q}^4 \rho _{y_q} }{r_q^4}\right]\psi ''\left(r_q\right)\\
    &+\frac{ w_q \left(x-x_q\right)^2 \left(y-y_q\right)^2  \rho _{x_q}^4 \rho _{y_q}}{r_q^3}\psi'''\left(r_q\right)\\
\end{aligned}
\end{equation*}

\subsubsection*{Partial derivatives along \textit{y}}

\begin{equation*}
     \partial_y\phi=\sum_{q=1}^Nc_q \psi \left(r_q\right)+\frac{w_q \left(y-y_q\right) \rho _{y_q}^2 }{r_q}\psi '\left(r_q\right)
\end{equation*}
\begin{equation*}
     \partial_{yy}\phi=\sum_{q=1}^N
     \left[
     \frac{\left(w_q+c_q \left(y-y_q\right)\right) \rho _{y_q}^2 }{r_q}
     -\frac{w_q \left(y-y_q\right)^2 \rho _{y_q}^4 }{r_q^3}
     \right]\psi '\left(r_q\right)
     +\frac{w_q \left(y-y_q\right)^2 \rho _{y_q}^4 }{r_q^2}\psi ''\left(r_q\right)
\end{equation*}
\begin{equation*}
    \partial_{ya_q}\phi=\frac{ \left(y-y_q\right) \rho _{y_q}^2}{r_q} \psi '\left(r_q\right)
\end{equation*}
\begin{equation*}
    \partial_{yya_q}\phi=\left[
    \frac{\rho _{y_q}^2 }{r_q}
    -\frac{\left(y-y_q\right){}^2 \rho _{y_q}^4 }{r_q^3}
    \right]\psi '\left(r_q\right)
    +\frac{\left(y-y_q\right){}^2 \rho _{y_q}^4 }{r_q^2}\psi ''\left(r_q\right)
\end{equation*}
\begin{equation*}
    \partial_{yb_q}\phi=\frac{\left(x-x_q\right) \left(y-y_q\right) \rho _{y_q}^2 }{r_q}\psi '\left(r_q\right)
\end{equation*}
\begin{equation*}
    \partial_{yyb_q}\phi=\left[
    \frac{\left(x-x_q\right) \rho _{y_q}^2 }{r_q}
    -\frac{\left(x-x_q\right) \left(y-y_q\right){}^2 \rho _{y_q}^4 }{r_q^3}
    \right]\psi '\left(r_q\right)
    +\frac{\left(x-x_q\right) \left(y-y_q\right){}^2 \rho _{y_q}^4 }{r_q^2}\psi ''\left(r_q\right)
\end{equation*}
\begin{equation*}
    \partial_{yc_q}\phi=\psi \left(r_q\right)+\frac{\left(y-y_q\right){}^2 \rho _{y_q}^2 }{r_q}\psi '\left(r_q\right)
\end{equation*}
\begin{equation*}
    \partial_{yyc_q}\phi=
    \left[\frac{3 \left(y-y_q\right) \rho _{y_q}^2}{r_q}
    -\frac{\left(y-y_q\right){}^3 \rho _{y_q}^4 }{r_q^3}\right] \psi '\left(r_q\right)
    +\frac{\left(y-y_q\right){}^3 \rho _{y_q}^4 }{r_q^2}\psi ''\left(r_q\right)
\end{equation*}
\begin{equation*}
    \partial_{y\rho_{x_q}}\phi=\left[
    \frac{c_q \left(x-x_q\right){}^2 \rho _{x_q} }{r_q}
    -\frac{w_q \left(x-x_q\right){}^2 \left(y-y_q\right) \rho _{x_q} \rho _{y_q}^2 }{r_q^3}\right]\psi '\left(r_q\right)
    +\frac{w_q \left(x-x_q\right){}^2 \left(y-y_q\right) \rho _{x_q} \rho _{y_q}^2 }{r_q^2}\psi ''\left(r_q\right)
\end{equation*}
\begin{equation*}
\begin{aligned}
    \partial_{yy\rho_{x_q}}\phi=&-\left[
    \frac{\left(w_q+2c_q \left(y-y_q\right)\right) \left(x-x_q\right){}^2 \rho _{x_q} \rho _{y_q}^2 }{r_q^3}
    -\frac{3w_q \left(x-x_q\right){}^2 \left(y-y_q\right){}^2 \rho _{x_q} \rho _{y_q}^4 }{r_q^5}\right]\psi '\left(r_q\right)\\
    &+\left[\frac{\left(w_q+2c_q \left(y-y_q\right) \right)\left(x-x_q\right){}^2 \rho _{x_q} \rho _{y_q}^2 }{r_q^2}
    -\frac{3w_q \left(x-x_q\right){}^2 \left(y-y_q\right){}^2 \rho _{x_q} \rho _{y_q}^4 }{r_q^4}
    \right]\psi ''\left(r_q\right)\\
    &+\frac{w_q \left(x-x_q\right){}^2 \left(y-y_q\right){}^2  \rho _{x_q} \rho _{y_q}^4}{r_q^3}\psi'''\left(r_q\right)\\
\end{aligned}
\end{equation*}
\begin{equation*}
    \partial_{y\rho_{y_q}}\phi=\left[
    \frac{\left(2w_q+c_q \left(y-y_q\right)\right)\left(y-y_q\right) \rho _{y_q} }{r_q}
    -\frac{w_q \left(y-y_q\right){}^3 \rho _{y_q}^3 }{r_q^3}\right]\psi '\left(r_q\right)
    +\frac{w_q \left(y-y_q\right){}^3 \rho _{y_q}^3 }{r_q^2}\psi ''\left(r_q\right)
\end{equation*}

\begin{equation*}
\begin{aligned}
    \partial_{yy\rho_{y_q}}\phi=&-\left[
    -\frac{2\left(w_q+2c_q \left(y-y_q\right)\right) \rho _{y_q} }{r_q}
    +\frac{\left(5w_q+2c_q \left(y-y_q\right)\right)\left(y-y_q\right) \rho _{y_q}^3 }{r_q^3}
    -\frac{3w_q \left(y-y_q\right){}^4 \rho _{y_q}^5 }{r_q^5}
    \right]\psi '\left(r_q\right)\\
    &+\left[\frac{\left(5w_q+2c_q \left(y-y_q\right)\right)\left(y-y_q\right)^2\rho _{y_q}^3 }{r_q^2}
    -\frac{3w_q \left(y-y_q\right){}^4 \rho _{y_q}^5 }{r_q^4}\right]\psi ''\left(r_q\right)\\
    &+\frac{w_q \left(y-y_q\right){}^4 \rho _{y_q}^5}{r_q^3}\psi'''\left(r_q\right)\\
\end{aligned}
\end{equation*}



\subsection*{A note about spatial dimensions}
 
The neural network can be extended to three dimensional phase unwrapping simply by adding the derivatives along the third dimension:
\begin{equation}
    \begin{matrix}
    &i_3=\cos(\partial_{xL}\phi_W)&i_9=\cos(\partial_{xR}\phi_W)\\
    &i_4=\sin(\partial_{xL}\phi_W)&i_{10}=\sin(\partial_{xR}\phi_W)\\
    i_1=\cos(\phi_W)&i_5=\cos(\partial_{yL}\phi_W)&i_{11}=\cos(\partial_{yR}\phi_W)&i_{15}=W(\partial_{xx}\phi_W)\\
    i_2=\sin(\phi_W)&i_6=\sin(\partial_{yL}\phi_W)&i_{12}=\sin(\partial_{yR}\phi_W)&i_{16}=W(\partial_{yy}\phi_W)\\
    &i_7=\cos(\partial_{zL}\phi_W)&i_{13}=\cos(\partial_{zR}\phi_W)&i_{17}=W(\partial_{zz}\phi_W)\\
    &i_8=\sin(\partial_{zL}\phi_W)&i_{14}=\sin(\partial_{zR}\phi_W)\\
    \end{matrix}
    \label{eq:inlayer_3D}
\end{equation}
with the output layer given by $\phi(x,y,z)=\sum_{q=1}^N w_q\psi \left(r_q\right)$ where
$r_q=\sqrt{(x-x_q)^2\rho^2_{x_q}+(y-y_q)^2\rho^2_{y_q}+(z-z_q)^2\rho^2_{z_q}}$
and $w_q=a_q+b_q(x-x_q)+c_q(y-y_q)+d_q(z-z_q)$.
Using the formulas from Eq. \ref{eq:inlayer_3D}, we get the corresponding output values $o_1,\hdots,o_{17}$. Neural networks with even higher number of dimensions can be built trivially by extending this procedure as necessary.

\bibliography{biblio}

\section*{Acknowledgements}

This research was supported by the NSF CAREER Award PHY-1943939.

\section*{Author contributions statement}

P.-A.G. developed the neural network architecture. He also tested the accuracy of the neural network on random phases and plasma interferograms. 
A.B. tested the accuracy of the neural network for quasi-monotonic phases

\end{document}